\documentclass[amssymb,
amsmath,
amsfonts,
reprint,
superscriptaddress,
longbibliography,
physrev, 
nofootinbib,
aps]{revtex4-2}

\usepackage[utf8]{inputenc}
\usepackage{graphicx}
\usepackage{dcolumn}
\usepackage[T1]{fontenc}
\usepackage{tabularx}
\usepackage{enumitem}
\usepackage{filecontents}
\usepackage{soul}
\usepackage{booktabs}
\usepackage{multirow}

\let\svthefootnote\thefootnote
\newcommand\freefootnote[1]{%
  \let\thefootnote\relax%
  \footnotetext{#1}%
  \let\thefootnote\svthefootnote%
}

\newif\iffastcompile
\fastcompilefalse

\usepackage[mathlines]{lineno}
\usepackage{graphicx}
\usepackage[margin=1in]{geometry}
\usepackage{graphicx}
\usepackage[table]{xcolor}
\usepackage{xcolor}

\usepackage[T1]{fontenc}
\usepackage{lmodern}
\usepackage{graphicx}
\usepackage{booktabs}
\usepackage{microtype}
\usepackage{placeins}

\newcommand{\pb}{\underline{p}}

\usepackage[noend]{algpseudocode}
\makeatletter
\newcounter{algorithm}
\def\fps@algorithm{tbp}
\def\ftype@algorithm{4}
\def\ext@algorithm{loa}
\def\fnum@algorithm{Algorithm~\thealgorithm}
\newenvironment{algorithm}[1][tbp]%
  {\@float{algorithm}[#1]\hrule\vspace{3pt}}%
  {\vspace{3pt}\hrule\end@float}
\makeatother

\usepackage[hidelinks, hypertexnames=false]{hyperref}
\definecolor{color1}{rgb}{0,0.25,0.70}
\hypersetup{colorlinks=true,
    linkcolor={color1},
    citecolor={color1},
    urlcolor={color1}
}

\newcommand{\ket}[1]{| #1 \rangle}

\begin{document}

\title{General Noise Channels and Mid-Circuit Measurements for Pre-Trajectory Sampling with Batched Execution}

\author{Louis Follet$^\dagger$}
\affiliation{NVIDIA Corporation, Santa Clara, CA, USA}
\affiliation{Research Laboratory of Electronics, Massachusetts Institute of Technology, Cambridge, MA 02139, USA}
\author{Paavai Pari}
\affiliation{NVIDIA Corporation, Santa Clara, CA, USA}
\author{Isidor Schoch}
\affiliation{NVIDIA Corporation, Santa Clara, CA, USA}
\affiliation{Department of Materials, University of Oxford, Parks Road, Oxford OX1 3PH, UK}
\author{Thien Nguyen}
\affiliation{NVIDIA Corporation, Santa Clara, CA, USA}
\author{Justin G. Lietz}
\affiliation{NVIDIA Corporation, Santa Clara, CA, USA}
\author{Daniel Lowell}
\affiliation{NVIDIA Corporation, Santa Clara, CA, USA}
\author{Brucek Khailany}
\affiliation{NVIDIA Corporation, Santa Clara, CA, USA}
\author{Krysta M. Svore}
\affiliation{NVIDIA Corporation, Santa Clara, CA, USA}
\author{Dirk R. Englund}
\affiliation{Research Laboratory of Electronics, Massachusetts Institute of Technology, Cambridge, MA 02139, USA}
\author{Taylor L. Patti$^\ddagger$}
\affiliation{NVIDIA Corporation, Santa Clara, CA, USA}

\freefootnote{
$^\dagger$louisf13@mit.edu, \\
$^\ddagger$tpatti@nvidia.com. 
}

\date{November 18, 2026}
\begin{abstract}

Many studies of quantum protocols require fast sampling of large amounts of measurement data from noisy quantum circuits, such as for training machine-learning decoders for quantum error correction. Exact simulation of realistic noise is expensive, while stabilizer-based simulators are fast but approximate, for example through Pauli twirling, which generally alters the output statistics. Pre-trajectory sampling with batched execution (PTSBE) accelerates exact trajectory simulation by sampling noise before state evolution. While very efficient, this method requires the outcome probability of every stochastic event to be independent of the state, which excludes the analysis of systems with mid-circuit measurements and general (non-mixture-of-unitaries) Kraus channels. We present in this work a generalization of PTSBE to measurement and Kraus operators without additional approximations. Each stochastic event is partially determined in advance: a state-independent part is pre-sampled, and the rest is drawn from the intermediate state. Trajectories with identical pre-sampled outcomes evolve as one group, a single multinomial draw splits the group at each deferred event, and branches resume from cached states. We implement the method for GPUs as an extension of CUDA-Q. On distance-3 topological code memories under amplitude damping, our implementation is up to $24 \times$ faster than conventional trajectory sampling in CUDA-Q. On 30-qubit random circuits with mid-circuit measurements, it is up to $83 \times$ faster and outperforms the near-Clifford simulator Clifft when more than about 25\% of the gates are non-Clifford.
With repeated sampling of each final state, its sampling rate per trajectory exceeds that of CUDA-Q by nearly five orders of magnitude. Our method also allows us to compare exact amplitude damping with its Pauli-twirled channel on a surface-code memory, revealing that twirling overestimates the detection rate by 21\% and raises the logical error rate of a lookup decoder by 8\%. 
This work makes exact, high-throughput simulation practical for general noisy circuits, from data generation for training decoders to studying measurement-based fault-tolerant protocols under realistic noise models.
\end{abstract}

\maketitle
\makeatletter
\let\si@origaddtocontents\addtocontents
\renewcommand{\addtocontents}[2]{%
  \def\si@tmp{#1}\def\si@toc{toc}%
  \ifx\si@tmp\si@toc\else\si@origaddtocontents{#1}{#2}\fi}
\makeatother

\clearpage

\section{Introduction}

The classical simulation of noisy quantum circuits serves many purposes, from predicting hardware behavior to generating training data for machine-learning models. Moreover, most such applications require large numbers of sampled measurement records, one per execution of the circuit, called a shot, under a hardware-accurate noise model. A prominent example is quantum error correction (QEC), which relies on repeated measurements to identify errors while preserving encoded quantum information \cite{fowler2012,terhal2015}. These measurement records are increasingly processed by machine-learning decoders \cite{torlai2017,bausch2024,alexeev2024}, which must be trained on large datasets representative of the noise encountered during computation. Hardware can provide such data directly, but collecting data at scale is costly and offers limited control over the underlying error processes. Simulation is a complementary data source, provided that it combines high sampling throughput with a faithful treatment of the noise model of interest.

The fastest QEC simulators achieve much of their efficiency by exploiting structure in the circuit or noise model. Stabilizer simulators, such as Stim~\cite{gidney2021stim}, enable extremely fast sampling of Clifford circuits with Pauli noise, while more recent methods such as Clifft~\cite{clifft} extend efficient simulation into the near-Clifford regime. More general noise processes can often be replaced by stochastic Pauli models, for example through Pauli twirling, making them compatible with such specialized simulators. This approximation often reproduces logical error rates well~\cite{gellerzhou2013, gutierrez2015}, but it can bias them~\cite{tomitasvore2014, katabarwa2015, garner2026} and does not in general preserve the full effect of the original channel on a QEC circuit. Approximations by Clifford channels and Pauli measurements come closer, but still do not reproduce the channel~\cite{magesan2013, puzzuoli2014, gutierrez2013}.

For applications in which simulated measurement records become the input to statistical analysis or machine-learning pipelines, changes in the distribution of these records can matter directly. Efforts to retain the exact channel have so far relied on density-matrix simulation, limited to about 17 qubits~\cite{obrien2017density}, or on tensor networks, which either assume perfect syndrome measurements~\cite{darmawan2017} or take seconds per shot~\cite{Barone_2026}. Quasiprobability decompositions also retain the exact channel and estimate circuit statistics without bias, but return signed-weight samples~\cite{pashayan2015, howard2017, Bennink2017, Seddon2019}.
Quantum-trajectory simulation  handles any channel through stochastic pure-state evolution~\cite{dalibard1992, dum1992, molmer1993, plenio1998, daley2014} and can be accelerated with a delayed inner product algorithm~\cite{isakov2021qsim}. Conventional trajectory simulators~\cite{cudaq, isakov2021qsim}, however, evolve every shot independently and closely related trajectories can repeat the same state evolution.

Several simulators exploit this redundancy. Approximate schemes reuse states across a fixed tree of subcircuits at a bounded cost in accuracy~\cite{wang2025tqsim}. Among exact simulators, some fix every noise outcome before the simulation starts, so that identical prefixes can be found and evolved once~\cite{li2019sanq, li2020redundant, dangwal2025tusq}. Pre-trajectory sampling with batched execution (PTSBE)~\cite{patti2025ptsbe, patti2026tn} also fixes all outcomes in advance, and gains its speed by evolving each distinct trajectory once and drawing its measurement records in a single batch, with control over how many records each trajectory produces, including nonproportional allocations. Fixing outcomes in advance, however, requires their probabilities to be independent of the state, which excludes mid-circuit measurements (MCMs) and general (non-mixture-of-unitaries) Kraus channels unless decoherence is first twirled into Pauli noise~\cite{dangwal2025tusq}. Other exact simulators draw each outcome during state evolution and copy the state shared by several shots only when their outcomes differ~\cite{doi2023shotbranching}, which handles any channel or measurement but fixes nothing in advance. To our knowledge, no exact simulator combines the speed of pre-sampled, batched execution with support for state-dependent operations.

\begin{figure*}[!t]
    \centering
    \includegraphics[width=1\textwidth]{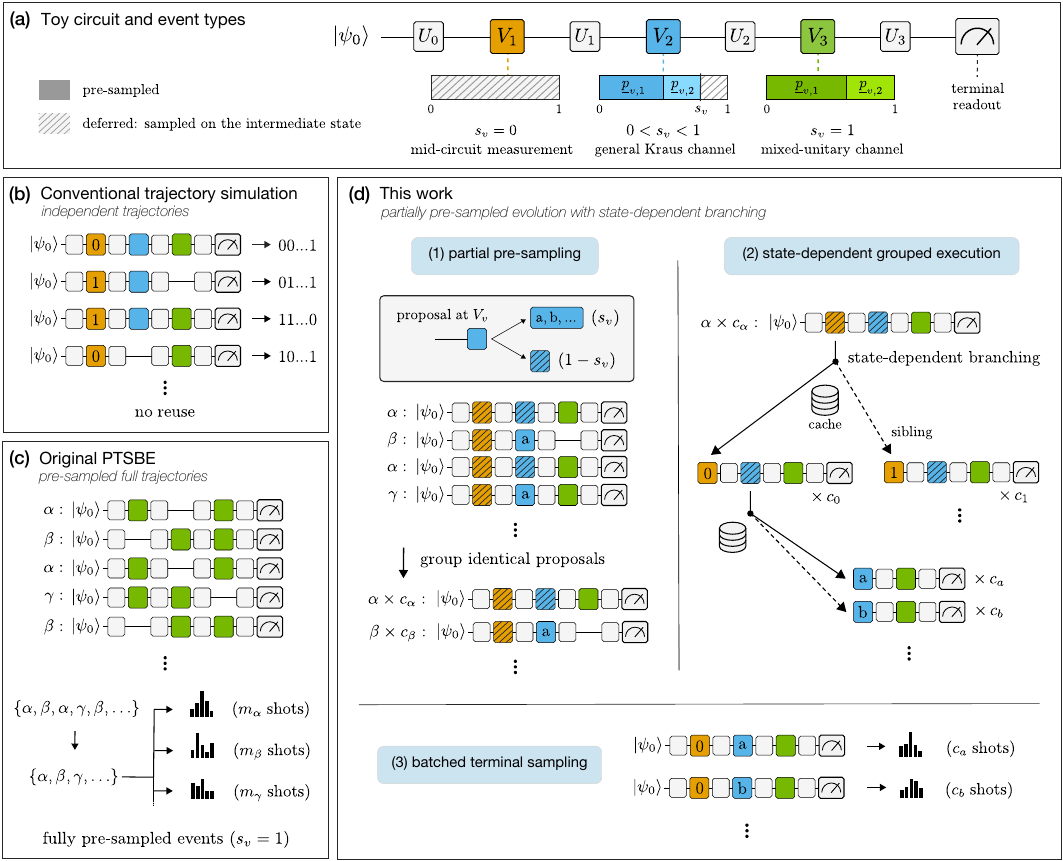}
\caption{\textbf{Pre-trajectory sampling with batched execution for state-dependent stochastic events.}
    \textbf{(a)} An example circuit with deterministic unitary blocks $U_i$ and stochastic events $V_v$, colored by their pre-sampling probability $s_v$ in all panels: orange for mid-circuit measurements ($s_v=0$), blue for general Kraus channels ($0<s_v<1$), and green for mixed-unitary channels ($s_v=1$). Each bar shows how the draw of an event splits into pre-sampled outcomes and a part drawn on the intermediate state (Sec.~\ref{sec:bg-dip}).
    \textbf{(b)} Conventional trajectory simulation draws stochastic outcomes during propagation and evolves each trajectory independently.
    \textbf{(c)} Original PTSBE pre-samples complete trajectories when every event has $s_v = 1$, deduplicates repeated trajectories, and prepares each retained trajectory $\alpha$ once for $m_\alpha$ terminal samples.
    \textbf{(d)} A diagram of the generalized PTSBE process presented in this work. (1)~At event $V_v$, a proposal stores either a pre-sampled outcome, with total probability $s_v$ (solid), or, with probability $1-s_v$, the residual distribution $\boldsymbol p^{\mathrm{res}}_v$ (hatched). Identical proposals are grouped with their multiplicity. (2)~At a deferred event, $\boldsymbol p^{\mathrm{res}}_v$ is evaluated once on the group's intermediate state, and a multinomial draw splits the group into sibling groups, one per occupied outcome. Dashed arrows mark siblings that resume from the cached state. (3)~Each trajectory is prepared once and sampled in a batch at the terminal readout, here with proportional sampling, such that each trajectory receives as many terminal samples as its multiplicity.}
    \label{Fig:overview}
\end{figure*}

In this work, we generalize PTSBE to circuits containing mid-circuit measurements and general Kraus channels [Fig.~\ref{Fig:overview}], in a GPU simulator based on CUDA-Q~\cite{cudaq}. 
With the lower bounds of the delayed inner product algorithm~\cite{isakov2021qsim}, the outcome probabilities of each stochastic event are split into a state-independent part, pre-sampled before execution, and a residual distribution, evaluated on the intermediate state when the trajectory reaches the event. Each trajectory is thus first drawn as a proposal, which holds for every event either a pre-sampled outcome or, for a deferred event, the residual distribution. Identical proposals evolve as one group. At a deferred event, a single multinomial draw from the residual distribution splits the group that reaches it, and sibling groups restart from the state cached before the event, in a tree-based simulation structure~\cite{li2020redundant,wang2025tqsim}.

Together, these mechanisms retain the computational advantages of PTSBE across state-dependent operations and without approximation. On distance-3 topological code memories under amplitude damping, generalized PTSBE achieves 11--24$\times$ higher sampling throughput than conventional trajectory sampling with CUDA-Q v0.15.1, a leading GPU-accelerated statevector simulator~\cite{cudaq, cudaqnoisy}. 
On random monitored circuits of 16 to 30 qubits with Pauli noise, it is 47--318$\times$ faster than CUDA-Q, and it outperforms Clifft above a crossover density of non-Clifford gates that falls from 62\% to 24\% as the qubit count grows, reaching up to a $10^3\times$ speedup. With amplitude damping, which Clifft does not support, the speedup over CUDA-Q is 83--176$\times$.
In the nonproportional sampling regime, our sampling rate per trajectory exceeds that of CUDA-Q by a factor of more than $7 \times 10^4$, and about 90\% of the shots are distinct in their trajectory and terminal bits.

The rest of this manuscript is organized as follows. Section~\ref{sec:background} reviews trajectory simulation, the original, fully pre-sampled PTSBE, and the delayed inner product algorithm. Section~\ref{sec:method} generalizes PTSBE to partially pre-sampled trajectories, with grouped execution and terminal readout. Section~\ref{sec:perf} first demonstrates that replacing amplitude damping by its Pauli-twirled channel changes the syndrome statistics of a surface-code memory, motivating exact channel simulation, and then benchmarks generalized PTSBE against CUDA-Q and Clifft. Finally, Section~\ref{sec:discussion} discusses the origin of the speedups, exact noise-model advantages, and future research directions.

\section{Background}
\label{sec:background}

The generalized implementation of PTSBE presented in this manuscript combines the trajectory reuse of the original PTSBE 
\cite{patti2025ptsbe, patti2026tn} with the state-independent lower bounds 
of the delayed inner product algorithm \cite{isakov2021qsim}. We review both
 after summarizing the foundations of conventional trajectory simulation for noisy quantum systems. 

\subsection{Quantum trajectories}
\label{sec:bg-trajectories}

We consider a circuit in which deterministic unitary blocks $U_i$ are 
interleaved with stochastic events $V_v$, $v=1,\dots,L$, such as noise channels and 
mid-circuit measurements [Fig.~\ref{Fig:overview}(a)]. Each trajectory is 
represented by a statevector initialized in $\ket{\psi_0}$. For event $V_v$, 
let $\mathcal A_v$ denote the finite set of its possible outcomes, such as which Kraus operator of a noise channel acts or which result a measurement gives, and 
$\{K_{v,a}\}_{a\in\mathcal A_v}$ the associated operators, satisfying
\begin{equation}
  \sum_{a\in\mathcal A_v} K_{v,a}^{\dagger}K_{v,a}=I.
  \label{eq:kraus-completeness}
\end{equation}

In a trajectory, a noise channel acting on density matrices as $\rho \mapsto \sum_{a}K_{v,a}\rho K_{v,a}^\dagger$ applies its Kraus operator $K_{v,a}$ when outcome $a$ occurs, and a projective measurement applies the projector $K_{v,a}$ of result $a$, for instance $K_{v,a}=|a\rangle\langle a|$ with $a\in\{0,1\}$ for a single-qubit measurement. 
For a normalized $\ket{\psi}$ immediately before an event, the probability
of outcome $a$ is 
\begin{equation}
  p_{v,a}(\psi)=\langle\psi|K_{v,a}^{\dagger}K_{v,a}|\psi\rangle,
  \label{eq:physical-outcome-probability}
\end{equation}
and, for $p_{v,a}(\psi) > 0$, the corresponding renormalized state is 
\begin{equation}
  |\psi_{v,a}\rangle
  =\frac{K_{v,a}|\psi\rangle}{\sqrt{p_{v,a}(\psi)}}.
  \label{eq:physical-state-update}
\end{equation}

For a mid-circuit measurement, outcome $a$ is also added to the classical measurement record. Trajectory simulation draws the outcome of each stochastic event with the probabilities of Eq.~\eqref{eq:physical-outcome-probability} on the intermediate state and updates the state by Eq.~\eqref{eq:physical-state-update}. Conventional implementations evolve one independent trajectory per shot [Fig.~\ref{Fig:overview}(b)].

\subsection{Original PTSBE}
\label{sec:bg-ptsbe}

In its original version, PTSBE \cite{patti2025ptsbe, patti2026tn} obtains its efficiency in two steps, illustrated in Fig.~\ref{Fig:overview}(c).
It first samples the noise outcomes of every trajectory before state evolution (pre-trajectory sampling) and groups repeated trajectories.
It then evolves each distinct trajectory once and reuses its final state for batched terminal readout (batched execution). Each final state is sampled either a fixed, possibly large, number of times or in proportion to the trajectory’s probability.
However, selecting outcomes in advance requires state-independent probabilities.
By Eq.~\eqref{eq:physical-outcome-probability}, this means that original PTSBE only works for operators conforming to $K_{v,a}^\dagger K_{v,a} \propto I$, i.e., Kraus operators that are proportional to unitaries, as is the case for Pauli and other 
mixture-of-unitary channels. As a result, MCMs, qubit resets, and general Kraus channels fall outside the scope of the original PTSBE algorithm.

\subsection{The delayed inner product algorithm}
\label{sec:bg-dip}

The delayed inner product algorithm~\cite{isakov2021qsim} assigns to each outcome $a$ of event $V_v$ a probability lower bound that holds for any input state,   
\begin{equation}
\underline p_{v,a}=\sigma_{\min}(K_{v,a})^2, \qquad s_v=\sum_{a\in\mathcal A_v}\underline p_{v,a},
\label{eq:bounds}
\end{equation}
where $\sigma_{\min}(K_{v,a})$ denotes the smallest singular value of $K_{v,a}$, so that $\sigma_{\min}(K_{v,a})^2$ is the smallest eigenvalue of $K_{v,a}^\dagger K_{v,a}$. The quantity $\underline p_{v,a}$ is the largest state-independent lower bound on $p_{v,a}(\psi)=\|K_{v,a}|\psi\rangle\|^2$ over all normalized input states, 
attained when $|\psi\rangle$ is an eigenvector for this eigenvalue. Since the total probability is normalized for any state, $0\le s_v\le1$.

At each event, the algorithm draws a number $r$ uniformly between 0 and 1 and compares it with $s_v$.
\begin{itemize}
    \item If $r<s_v$, the interval $[0,s_v)$ is cut into consecutive segments of length $\underline p_{v,a}$, one per outcome, which are known before the simulation. Outcome $a$ is selected if $r$ falls in its segment, without evaluating any $p_{v,a}(\psi)$.
    \item If $r\ge s_v$, the interval $[s_v,1)$ is cut into segments of length $p_{v,a}(\psi)-\underline p_{v,a}$, the excess of each probability over its bound. These lengths depend on the intermediate state, so they are computed only in this case.
\end{itemize} 
Outcome $a$ thus covers a total length $\underline p_{v,a} + [p_{v,a}(\psi)-\underline p_{v,a}] = p_{v,a}(\psi)$,
and the selection follows Eq.~\eqref{eq:physical-outcome-probability} exactly.
The bound sum $s_v$ is the probability that $r<s_v$, so that the outcome is selected without evaluating any probability on the state, and we call it the pre-sampling probability of the event.
The algorithm draws $r$ separately for each trajectory when it reaches the event. It therefore skips evaluating $p_{v,a}(\psi)=\langle\psi|K_{v,a}^\dagger K_{v,a}|\psi\rangle$, an inner product with the state, whenever $r<s_v$, but still evolves every trajectory separately.

\section{Generalized PTSBE}
\label{sec:method}

\subsection{Partially pre-sampled trajectories}
\label{sec:partially-resolved}

The original PTSBE requires $s_v=1$ at every event. We relax this requirement by allowing trajectories to remain partially pre-sampled until the relevant quantum state becomes available. 

For $s_v<1$, normalizing the state-dependent segments of Sec.~\ref{sec:bg-dip} by their total length $1-s_v$ gives the residual probability distribution
\begin{equation}
p^{\mathrm{res}}_{v,a}(\psi)=\frac{p_{v,a}(\psi)-\underline p_{v,a}}{1-s_v}.
\label{eq:residual}
\end{equation}
These probabilities are nonnegative and sum to one, giving the exact decomposition
\begin{equation}
p_{v,a}(\psi)=\underline p_{v,a}+(1-s_v)\,p^{\mathrm{res}}_{v,a}(\psi).
\label{eq:decomposition}
\end{equation}
This decomposition allows part of the outcome selection to take place before state evolution. Before execution, we draw for each event $V_v$ a random variable $Z_v$ with state-independent probabilities
\begin{equation}
\begin{aligned}
\Pr(Z_v=a) &= \underline p_{v,a} && (a\in\mathcal A_v),\\
\Pr\bigl(Z_v= \boldsymbol p^{\mathrm{res}}_v\bigr) &= 1-s_v && (s_v<1).
\end{aligned}
\label{eq:presampling}
\end{equation}
If $Z_v=a$, outcome $a$ is pre-sampled. If $Z_v= \boldsymbol p^{\mathrm{res}}_v$, the event is deferred, and its outcome is drawn at run time from the residual distribution evaluated on the intermediate state. By Eq.~\eqref{eq:decomposition}, the total probability of selecting $a$ is $p_{v,a}(\psi)$, with the corresponding state given by Eq.~\eqref{eq:physical-state-update}. A partially pre-sampled trajectory is represented by the proposal sequence
\begin{equation}
\boldsymbol Z=(Z_1,\ldots,Z_L),\qquad Z_v\in\mathcal A_v\cup\bigl\{\boldsymbol p^{\mathrm{res}}_v\bigr\}.
\label{eq:proposal}
\end{equation}

When $s_v=1$ at every event, all outcomes are selected before state evolution and the original PTSBE is recovered [Fig.~\ref{Fig:overview}(c)], which happens exactly for Kraus operators proportional to unitaries (Supplementary Note~\ref{app:maxs-bound}). Nontrivial MCMs have $s_v=0$, so their outcomes are always sampled during execution. General Kraus channels, including nonunital noise channels, can take intermediate values $0<s_v<1$, giving the partially pre-sampled events illustrated in Fig.~\ref{Fig:overview}(a).

Amplitude damping, for instance, has Kraus operators $K_{v,0}=\operatorname{diag}(1,\sqrt{1-\gamma})$ for no decay and $K_{v,1}=\sqrt{\gamma}\,|0\rangle\langle1|$ for decay, where $\gamma$ is the decay probability of $|1\rangle$. By Eq.~\eqref{eq:bounds}, $\underline p_{v,0}=1-\gamma$ and $\underline p_{v,1}=0$, so a trajectory needs the state at a damping location only with probability $\gamma$.

\subsection{Grouped execution and terminal sampling}\label{sec:grouped}
Algorithm~\ref{alg:ptsbe} summarizes generalized PTSBE. It consists of a preprocessing step and the three stages of Fig.~\ref{Fig:overview}(d), which the paragraphs below motivate in order. Supplementary Note~\ref{app:exactness} proves that the trajectories follow the law of $N$ independent trajectory simulations.\\

\begin{algorithm}[tb]
\caption{Generalized PTSBE. \textsc{Execute}$(\boldsymbol Z,c,|\psi\rangle,v_0)$ evolves $c$ identical proposals $\boldsymbol Z$ from $|\psi\rangle$, starting at event $V_{v_0}$.}
\label{alg:ptsbe}
\small
\begin{algorithmic}[1]
\Require $\{U_i\}$, $\{V_v\}$, $|\psi_0\rangle$, $N$, mode for $m_\alpha$
\Statex \textbf{Preprocessing}
\State $\underline p_{v,a},\,s_v\ \forall v$ \Comment{Eq.~\eqref{eq:bounds}}
\Statex \textbf{Stage 1: pre-sampling and grouping}
\State $Z^{(n)}_v\sim$ Eq.~\eqref{eq:presampling}, independently $\forall n\le N,\ \forall v$
\State Group identical $\boldsymbol Z^{(n)}$ into groups $(\boldsymbol Z_g,c_g)$
\Statex \textbf{Stage 2: grouped execution, depth first}
\ForAll{groups $g$} \Call{Execute}{$\boldsymbol Z_g,c_g,|\psi_0\rangle,1$} \EndFor
\Statex \textbf{Stage 3: terminal sampling}
\ForAll{$\alpha$} draw $m_\alpha$ samples from $|\psi_\alpha\rangle$ \EndFor
\Statex
\Procedure{Execute}{$\boldsymbol Z,c,|\psi\rangle,v_0$}
  \For{$v=v_0,\ldots,L$}
    \State $|\psi\rangle\gets U_{v-1}|\psi\rangle$
    \If{$Z_v=\boldsymbol p^{\mathrm{res}}_v$} \Comment{deferred}
            \State $(C_a)_{a\in\mathcal A_v}\sim\mathrm{Multinomial}\bigl(c,\boldsymbol p^{\mathrm{res}}_v(\psi)\bigr)$
      \Statex \Comment{Eq.~\eqref{eq:multinomial}}
      \State Cache $|\psi\rangle$ if several $C_a>0$
      \ForAll{$a$ with $C_a>0$} 
        \State $Z_v\gets a$
        \State \Call{Execute}{$\boldsymbol Z,C_a,|\psi_{v,a}\rangle,v+1$}
      \EndFor
      \State \Return
    \EndIf
    \State $a\gets Z_v$,\ \ $|\psi\rangle\gets|\psi_{v,a}\rangle$ \Comment{Eq.~\eqref{eq:physical-state-update}}
  \EndFor
  \State $\alpha\gets\boldsymbol Z$ \Comment{all outcomes fixed: trajectory $\alpha$}
  \State $|\psi_\alpha\rangle\gets U_L|\psi\rangle$ \Comment{final state of $\alpha$}
  \State $c_\alpha\gets c_\alpha+c$ \Comment{multiplicity of $\alpha$}
\EndProcedure
\end{algorithmic}
\end{algorithm}

\textit{Preprocessing.} Once per channel and before constructing any states, we compute the lower bounds and pre-sampling probabilities of Eq.~\eqref{eq:bounds} and the operators that evaluate Eq.~\eqref{eq:residual}. 
The Kraus representation of a noise channel is not unique, and $s_v$ depends on it. Refactoring the Kraus operators to raise $s_v$ reduces the expected number of deferred events per proposal, $\sum_v(1-s_v)$, and we provide a search over equivalent representations that maximizes it (Supplementary Note~\ref{app:maxs}). The standard form of amplitude damping, used later in our benchmarks, is already optimal. A larger $s_v$ does not guarantee a shorter runtime, because the new representation also changes how often trajectories coincide and branch, so we propose this search as an optional preprocessing step for other noise models.\\

\textit{Pre-sampling and grouping.} Using only the lower bounds, we independently sample $N$ proposals, Eq.~\eqref{eq:proposal}, from Eq.~\eqref{eq:presampling} and group identical proposals with their multiplicities [Fig.~\ref{Fig:overview}(d), stage 1]. Each distinct proposal is then evolved once for all its copies. 
A group $g$ contains $c_g$ identical proposals $\boldsymbol Z_g$, which have the same pre-sampled outcomes and the same deferred events. These proposals share a quantum state $|\psi_g\rangle$ and advance through the circuit together.\\

\textit{Pre-sampled and deferred events.} When a group reaches the stochastic event $V_v$, it proceeds based on its associated random variable $Z_v$ [Fig.~\ref{Fig:overview}(d), stage 2]. If $Z_v=a$, all $c_g$ proposals advance through outcome $a$ by Eq.~\eqref{eq:physical-state-update}, without evaluating the residual distribution. If $Z_v=\boldsymbol p^{\mathrm{res}}_v$, the residual distribution is evaluated a single time on the shared state $|\psi_g\rangle$, for the whole group. The group is then split by a single multinomial draw
\begin{equation}
\boldsymbol C_g\sim\operatorname{Multinomial}\bigl(c_g,\boldsymbol p^{\mathrm{res}}_v(\psi_g)\bigr),
\label{eq:multinomial}
\end{equation}
where $C_{g,a}$ is the number of proposals assigned to outcome $a$. This draw is exact because the members share $|\psi_g\rangle$, so their residual draws are independent and identically distributed. Outcomes with $C_{g,a}=0$ are never simulated. For each outcome with $C_{g,a}>0$, applying $K_{v,a}$ according to Eq.~\eqref{eq:physical-state-update} produces a successor group of $C_{g,a}$ proposals. Each successor group then continues, updating the measurement record at mid-circuit measurements.\\

\textit{Branching and caching.} When the multinomial draw gives several outcomes with $C_{g,a}>0$, the group splits into sibling groups, one per such outcome, which all start from the state cached immediately before the event.
The first branch continues from the intermediate state, and the others restart from the cached copy. Depth-first traversal bounds the number of cached states by the number of deferred events on the current path (Supplementary Note~\ref{app:cache}). Different proposals can produce the same trajectory, because an outcome can be pre-sampled or drawn on the intermediate state. Their multiplicities are therefore added.\\

\textit{Terminal sampling.} Let $c_\alpha$ denote the accumulated multiplicity of a trajectory $\alpha$. Each trajectory is re-executed once to prepare its final state, from which $m_\alpha$ measurement records are drawn in a batch [Fig.~\ref{Fig:overview}(d), stage 3].
We provide three modes for $m_\alpha$: 1) proportional sampling, $m_\alpha=c_\alpha$, which gives the same distribution of measurement counts as independent trajectory sampling, 2) rescaled proportional sampling, $m_\alpha=b\,c_\alpha$ with a positive integer $b$, which draws more records from the same sampled trajectories, and 3) nonproportional sampling, $m_\alpha=m$ for every trajectory independently of its multiplicity (suited to generating large labeled datasets), with the terminal samples of each trajectory weighted by $c_\alpha/N$ to estimate circuit statistics (Supplementary Note~\ref{app:sampling}).

\begin{figure*}[!t]
    \centering
    \includegraphics[width=1\textwidth]{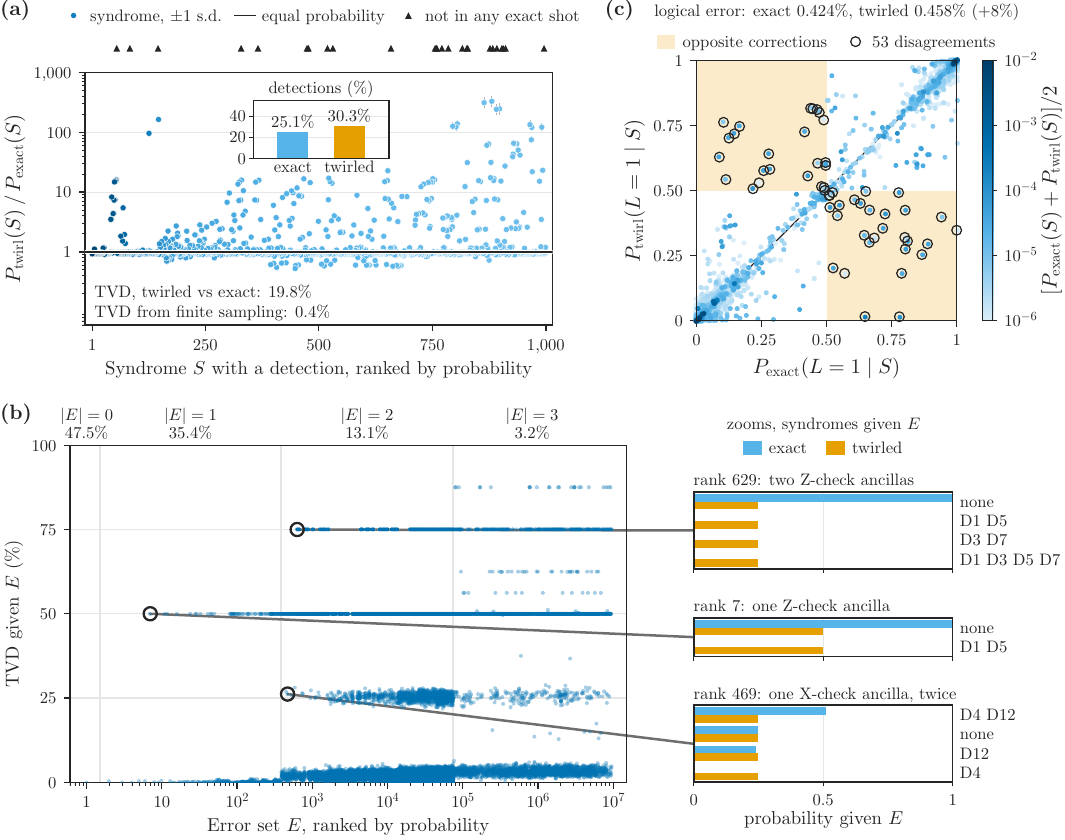}
    \caption{\textbf{Effect of Pauli twirling on syndrome statistics and decoding.} Exact amplitude damping (generalized PTSBE) is compared with its twirled channel (Stim) for a distance-3 rotated surface-code memory storing $|0\rangle_L$ over three rounds. In (a) and (c), points are shaded by $\tfrac12[P_{\mathrm{exact}}(S)+P_{\mathrm{twirl}}(S)]$.
\textbf{(a)}~Ratio $P_{\mathrm{twirl}}(S)/P_{\mathrm{exact}}(S)$ for the 1,000 most probable syndromes with a detection, with $\pm1$ s.d. bars. Triangles mark syndromes absent from all exact shots. The TVD, Eq.~\eqref{eq:tvd}, is computed given a detection and compared with the TVD that finite sampling alone would produce. Inset: share of shots with at least one detection. Syndrome data sampled with the twirled channel, as stabilizer simulators require, therefore misweight individual syndromes by up to a factor of 329 and overestimate the detection rate by 21\%.
\textbf{(b)}~TVD between the exact and twirled syndrome distributions given the error set $E$, the locations where a decay can occur. Error sets are ordered from most to least probable, and vertical lines separate the regions with $|E|=0,1,2,3$ locations, each labeled with its share of shots. The distance grows with $|E|$ and clusters on horizontal plateaus at 25\%, 50\%, and 75\%. The insets explain these plateaus by breaking down the syndrome mismatch for three example error sets, with each syndrome listed by the detectors $D_i$ that fire. For instance, exact damping leaves a Z-check ancilla in $|0\rangle$ quiet, whereas the twirled channel erroneously fires its detector pair half the time (50\% plateau).
\textbf{(c)}~Logical-flip probability $P(L=1\mid S)$ of each syndrome under the exact and twirled channels. Circled syndromes lie on opposite sides of $1/2$ (tinted quadrants), so lookup decoders built from the two models apply opposite corrections to them. Decoding with twirled statistics raises the logical error rate from 0.424\% to 0.458\% on $10^7$ exact shots. Shown are syndromes seen in at least 100 shots under both models (99.75\% of exact shots, standard errors at most 0.05).}
    \label{Fig:Fig2}
\end{figure*}

\begin{figure*}[!t]
    \centering
    \includegraphics[width=1\textwidth]{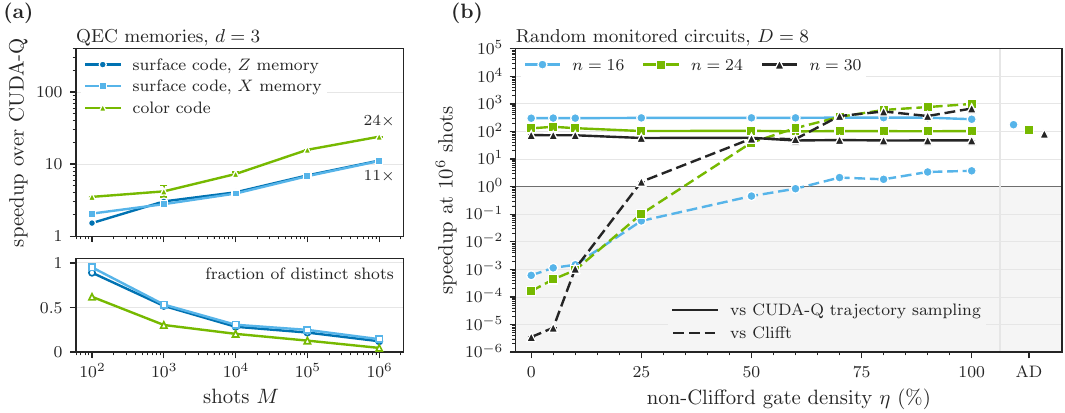}
    \caption{\textbf{Speedup of generalized PTSBE over trajectory sampling and near-Clifford simulation.}
       \textbf{(a)} Speedup over trajectory sampling in CUDA-Q (top) and fraction of distinct shots, the number of distinct (trajectory, terminal bitstring) pairs per shot (bottom), against the number of shots $M$, for three distance-3 memories run for three rounds under amplitude damping. Clifft is not shown, since it can only approximate this noise channel.
      \textbf{(b)} Speedup at $10^6$ shots over CUDA-Q (solid) and over Clifft (dashed), against the density $\eta$ of non-Clifford gates in random monitored circuits of $n=16$, 24, and 30 qubits and depth $D=8$. Each Clifford gate is replaced with probability $\eta$ by a non-Clifford Pauli rotation, under the same noise at every $\eta$. Generalized PTSBE is faster than CUDA-Q at every density and outperforms Clifft beyond a crossover density that falls from 62\% at $n=16$ to 24\% at $n=30$. Clifft is faster in the shaded region, where the speedup is below 1. The points labeled AD repeat the $\eta=100\%$ circuits with amplitude damping in place of the Pauli gate noise, which Clifft does not support. Points are geometric means over three runs, and bars show one standard deviation of the log rate.}
    \label{Fig:Fig3}
\end{figure*}

\section{Applications and performance}
\label{sec:perf}

We generalize PTSBE to trajectory sampling with any Kraus channel and mid-circuit measurement, for general-purpose circuit simulation as well as QEC dataset generation. We first show why exact simulation of such channels matters, by comparing amplitude damping with its Pauli-twirled channel on a surface-code memory (Sec.~\ref{sec:qec}). We then measure the speedup on random monitored circuits and QEC memories over exact conventional trajectory sampling and over the near-Clifford simulator Clifft in the noise regime it supports (Sec.~\ref{sec:accelerate}).

\subsection{QEC statistics with general noise channels}\label{sec:qec}
We consider a distance-3 rotated surface-code memory over three rounds, with single-qubit amplitude damping after every layer as the only noise source (Supplementary Note~\ref{app:qec-circuit}). Each shot yields a syndrome $S$, the outcomes of the 24 detectors comparing successive stabilizer measurements, and a logical bit $L$, equal to 1 when the stored state is flipped. We compare two noise models, the exact channel, simulated with our method, and the Pauli-twirled channel, sampled with Stim~\cite{gidney2021stim}, through the total-variation distance (TVD) between their distributions of the syndrome $S$,
\begin{equation}
d_{\mathrm{TV}}=\frac12\sum_S\bigl\lvert P_{\mathrm{exact}}(S)-P_{\mathrm{twirl}}(S)\bigr\rvert,
\label{eq:tvd}
\end{equation}
which ranges from 0 for identical distributions to 100\% for distributions with no syndrome in common.

Given a detection, the two distributions lie at a TVD of 19.8\%, well above the 0.4\% that finite sampling alone produces with about $10^9$ sampled syndromes each.
This disagreement appears both in individual syndromes, whose probabilities change by up to 329-fold [Fig.~\ref{Fig:Fig2}(a), Supplementary Note~\ref{app:syndrome-statistics}], and in the detection rate, which the twirled noise overestimates by 21\%, with at least one detection in 30.3\% of shots against 25.1\% under exact noise [Fig.~\ref{Fig:Fig2}(a), inset].

To locate where the two models differ, we sort shots by their error set $E$ [Fig.~\ref{Fig:Fig2}(b)], the locations where a decay can occur. Our pre-sampling identifies these locations as the deferred events of a proposal. Each location enters $E$ with probability $\gamma_v$, so $P(E)$ is known in closed form and lets us estimate the exact distribution set by set (Supplementary Note~\ref{app:stratified-estimation}). We draw the Pauli errors of the twirled channel at the same locations and compare the two models for each $E$. The distance is near zero when $E$ is empty, since no decay can occur. It grows with $|E|$, because each location in $E$ is a place where the two models act differently. For each $|E|$, distances cluster near zero and along plateaus at 25\%, 50\%, and 75\%.
The syndrome distributions of three example error sets [Fig.~\ref{Fig:Fig2}(b), zooms] reveal the origin of this structure. Amplitude damping leaves $|0\rangle$ unchanged, whereas the twirled channel can flip it. Consequently, a location of $E$ on a Z-check ancilla in $|0\rangle$ produces no detection event under exact damping, but triggers a detector pair half the time under the twirled channel, generating the 50\% plateau. With two such locations, the Pauli twirling approximation spreads over four equally likely syndromes and the distance reaches 75\%.

Finally, a decoder built from these statistics inherits the identified mismatch [Fig.~\ref{Fig:Fig2}(c)]. We build a lookup table from each model and test both on the same $10^7$ exact-noise records. The table built from the twirled model fails on 0.458\% of these records, against 0.424\% for the exact table, an 8\% relative increase in logical error rate. Because the test records follow the exact noise, each syndrome counts in proportion to its exact probability.
Fig.~\ref{Fig:Fig2}(c) shows only syndromes seen in at least 100 shots under both models. Given that rarer syndromes arise mostly from larger error sets, where the models differ more [Fig.~\ref{Fig:Fig2}(b)], one could expect them to receive opposite corrections more often.
This lookup decoder is a simple case, which uses only the side of $1/2$ on which each estimate of $P(L=1\mid S)$ falls, and twirling could affect more sophisticated decoders more. For instance, soft-output decoders, which estimate this probability itself~\cite{meister2024}, also respond to shifts that do not cross $1/2$, and learned decoders that take the full measurement bitstrings as input~\cite{bausch2024}, not only the syndromes computed from them, are also sensitive to differences that the syndromes do not show.

\subsection{Accelerating exact trajectory simulation}
\label{sec:accelerate}

We compare the sampling throughput of our implementation with conventional trajectory sampling in CUDA-Q v0.15.1~\cite{cudaq}, using its cuStateVec-based GPU statevector simulator~\cite{cudaqnoisy}, in both terminal allocation modes of Sec.~\ref{sec:grouped}. We also compare with the near-Clifford simulator Clifft~\cite{clifft} on circuits with only Pauli noise, since it does not support noise channels with non-unitary Kraus operators, such as amplitude damping. Our generalized PTSBE and CUDA-Q both evolve single-precision GPU statevectors under the same exact channels on one GPU. Clifft runs on 16 CPU cores. We expect the timings reported here to improve as we continue to develop our open-source implementation (Supplementary Note~\ref{app:benchmark-timing}).

\subsubsection{Proportional sampling}

We first run the 17-qubit surface-code memory of Sec.~\ref{sec:qec}, its X-basis counterpart, and a 10-qubit distance-3 color-code memory, all under amplitude damping, and measure the throughput for $M$ shots, one per proposal [Fig.~\ref{Fig:Fig3}(a)]. At $M = 10^2$, the speedup is $1.5\times$ to $3.5 \times$, depending on the memory circuit, because the $10^2$ draws contain only 54 distinct proposals in each surface-code memory and 37 in the color code. As $M$ increases, the most probable trajectories keep recurring and the fraction of distinct shots falls, so that at $M = 10^6$ the speedup reaches $11 \times$ for both surface-code memories. The color code has fewer noise locations, 261 against 384, so its trajectories repeat more often, leading to a greater speedup of $24 \times$.

To go beyond QEC circuits, we consider random monitored circuits \cite{skinner2019, li2019} of $n = 16$, 24, and 30 qubits with $D = 8$ alternating layers of random single- and two-qubit Clifford gates [Fig.~\ref{Fig:Fig3}(b)]. Each gate is replaced with probability $\eta$ by a non-Clifford Pauli rotation, so that $\eta$ sets the non-Clifford gate density. Some qubits are measured and reset after every second layer, and Pauli gate noise of probability $10^{-3}$ acts throughout, independently of $\eta$. Details and $D = 16$ circuits are discussed in Supplementary Note~\ref{app:benchmarks}.
We sweep $\eta$ and compare with CUDA-Q and Clifft. Near the Clifford regime, Clifft is faster since it is designed for such circuits. Its cost grows with the number of non-Clifford rotations, however, and our method outperforms it above $\eta\approx62\%$, 35\%, and 24\% for $n=16$, 24, and 30, reaching up to $3.8\times$, $10^3\times$, and $681\times$ respectively at $\eta=1$.
The speedup over CUDA-Q is nearly flat in $\eta$, reaching 47--318$\times$ depending on the circuit size, since a statevector engine applies Clifford and non-Clifford gates at the same cost.

We also run the $\eta = 1$ circuits with the Pauli gate noise replaced by amplitude damping (AD) of strength $\gamma = 10^{-3}$ after single-qubit layers and $5 \times 10^{-4}$ per qubit after two-qubit layers [Fig.~\ref{Fig:Fig3}(b), AD]. Stabilizer and near-Clifford simulators such as Stim and Clifft can only approximate this channel, so we compare with CUDA-Q only, giving speedups of 176$\times$, 111$\times$, and 83$\times$ for $n=16$, 24, and 30, of the same order as with Pauli noise.

The speedup decreases with $n$, because larger circuits have more noise locations, so trajectories repeat less often, and because the reported runs sample the final states only once all trajectories are complete, which requires preparing each of them again, a cost that grows with $n$. A future version of our implementation will offer an alternative readout mode, which samples each trajectory as soon as its final state is reached.

\subsubsection{Nonproportional sampling}

\begin{figure}[!t]
    \centering
    \includegraphics[width=1\columnwidth]{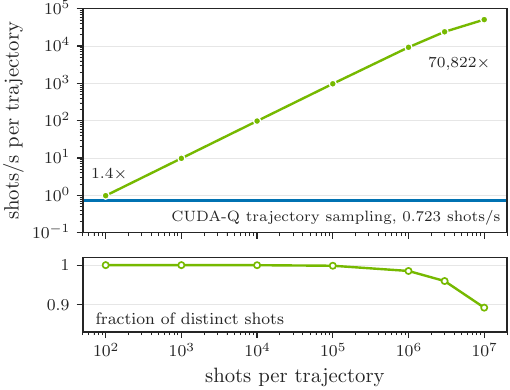}
    \caption{\textbf{Bulk generation of labeled shots with generalized PTSBE in nonproportional mode.}
A random monitored circuit of $n=30$ qubits and depth $D=16$, with amplitude damping on every qubit, is sampled from 100 trajectories that each receive the same number $m$ of shots. 
Top: shot rate per trajectory, the total rate divided by the number of trajectories, against $m$. Conventional trajectory sampling in CUDA-Q (blue) simulates one trajectory per shot. The speedup grows from $1.4 \times$ at $m=10^2$ to $70{,}000 \times$ at $m=10^7$. Bottom: fraction of distinct shots, the number of distinct (trajectory, terminal bitstring) pairs per shot. Each shot carries its trajectory label.}
    \label{Fig:Fig4}
\end{figure}

As in the original PTSBE algorithm \cite{patti2025ptsbe}, our simulator supports a nonproportional mode which assigns the same number $m$ of terminal samples to every trajectory, yielding many labeled shots per trajectory evolution, as needed for instance for large QEC datasets. We now apply it in the presence of state-dependent events, using a random monitored circuit with $n = 30$, $D = 16$, $\eta = 1$, and amplitude damping on every qubit, which records 14 mid-circuit and 30 terminal bits per shot [Fig.~\ref{Fig:Fig4}]. We draw 100 proposals, which give 100 distinct trajectories.

Conventional trajectory sampling in CUDA-Q evolves one trajectory per shot, at 0.723 shots/s, and we compare this rate with our shot rate per trajectory (Supplementary Note~\ref{app:benchmark-bulk}). The speedup rises from $1.4 \times$ at $m = 10^2$ to over $70{,}000 \times$ at $m = 10^7$, because generating the trajectories takes a fixed 46.3 s while sampling their final states grows only from 55.5 s to 149 s over five orders of magnitude in $m$. Across these batch sizes, the fraction of distinct shots stays above 0.89 [Fig.~\ref{Fig:Fig4}, bottom], so most additional terminal samples still yield new labeled records, for instance as training data for decoders.

All records share the noise realizations of these 100 trajectories, and each carries its trajectory label. Because the number of trajectories and the number of samples per trajectory are set independently, a dataset can cover more noise realizations or draw more terminal samples from each trajectory. When trajectories repeat, weighting each record by the multiplicity of its trajectory restores unbiased estimates of circuit statistics (Supplementary Note~\ref{app:flat-allocation}).

\section{Discussion}\label{sec:discussion}
Generalized PTSBE is faster than conventional trajectory sampling in all of the diverse benchmarks explored in this manuscript. In particular, it is up to $24\times$ faster on QEC memories in the proportional mode, up to 318$\times$ faster on random monitored circuits, and up to $70{,}000\times$ faster per trajectory in the nonproportional mode. It also outperforms the near-Clifford simulator Clifft beyond a crossover density of non-Clifford gates. In generalized PTSBE, each distinct trajectory is evolved just once, regardless of how many shots it serves, so the speedup depends on how many such trajectories and shots a circuit requires.

For QEC applications, these quantities grow with the number of likely error configurations, which is in turn set by code size, number of rounds, and error rates, such that larger codes and longer memories potentially lead to smaller speedups (Supplementary Note~\ref{app:cache}). Conversely, for larger codes, each evolution acts on a larger statevector, which increases the time saved per avoided evolution, as this grows exponentially in qubit number. The net effect of these two competing phenomena is system-dependent and could be evaluated in future studies. The nonproportional mode speedup is insensitive to the number of distinct trajectories, because its gain comes from repeated readout of each one, and thus strictly benefits from larger codes. Both modes should also benefit from further optimization of our implementation (Supplementary Note~\ref{app:benchmark-timing}).

Pauli twirling is often adopted so that high-throughput simulators can be used, but this choice is not generally accurate for QEC data. In our comparison, a lookup decoder built from twirled statistics makes 8\% more logical errors on exact noise, whereas the syndrome distribution that a decoder takes as input shifts by a TVD of 19.8\% given a detection. The conditional analysis traces this shift to error sets in which the twirled channel flips states that damping leaves unchanged. 
Stabilizer simulators can avoid the Pauli twirling approximation when a channel is a mixture of Clifford operations and reset with positive weights, as thermal relaxation is for $T_2\le T_1$~\cite{garner2026}. Pure amplitude damping, with $T_2=2T_1$ as in Sec.~\ref{sec:qec}, requires negative weights, which cannot serve as sampling probabilities, while our method samples it exactly.

Together, the speed and exactness of our approach enable large-scale simulation of protocols built on mid-circuit measurements and non-Clifford gates under realistic non-Pauli noise, providing a practical tool for the development of fault-tolerant quantum computers. In particular, at millions of labeled shots per second, each tagged with the trajectory that produced it, our method can supply the large and accurate datasets that AI-based decoders require under realistic nonunital channels, such as amplitude damping.

These capabilities open several directions for future work, as we expect the speedups of our method to benefit any study that samples many shots of a generally and realistically noisy circuit. For QEC, the most direct application is to train learned decoders on these extensive and exact datasets. Such decoders could also be compared with ones trained on twirled data, beyond the distance-three code and lookup decoder of Sec.~\ref{sec:qec}. The same contrast between exact and approximate noise models extends to logical gadgets, in particular to protocols based on repeat-until-success~\cite{paetznick2014, bocharov2015} or postselection, whose mid-circuit measurements our method treats exactly, such as gate synthesis, magic-state distillation~\cite{magicstatedistillation}, and cultivation~\cite{magicstatecultivationgrowing}. Identifying the regimes in which approximate models fail would accelerate the design of such gadgets and of their circuit implementations, a task that models trained on exact data could also support. Combining partial pre-sampling with tensor-network PTSBE, which has been demonstrated on circuits of up to 200 qubits with Pauli noise~\cite{patti2026tn}, could bring mid-circuit measurements and general noise channels beyond the 30 qubits of statevector simulation. Finally, our framework can be applied immediately to rare-event sampling, as is vital for achieving low logical error-rate decoders \cite{bravyi2013rare, heussen2024subset}.

\section*{acknowledgments}
L.F. performed part of this work during an internship at NVIDIA. L.F. and D.R.E. acknowledge research funding from NVIDIA to MIT. The authors acknowledge the use of AI tools to assist with code development and manuscript editing. The authors reviewed all AI-assisted output and take full responsibility for the content of this work.

\section*{code availability}
Our implementation will soon be released as open-source software.
\section{competing interests}
T.L.P., J.G.L., and L.F. are inventors on a patent application filed by NVIDIA, titled ``Statevector caching and batching techniques for efficient mid-circuit measurement simulations''. The other authors declare no competing interests.

\section*{Author contributions}
T.L.P., J.G.L., and L.F. conceived the project. L.F. developed the method and its software implementation, performed the simulations, and analyzed the results. I.S. contributed to the algorithm for general Kraus channels, T.N. to the software implementation, and J.G.L. to the quantum error correction applications. P.P. supported the final code development and optimization, large-scale benchmarks, and the preparation of the open-source software. D.L., B.K., and K.M.S. provided guidance on the applications, simulations, and analysis. T.L.P. and D.R.E. supervised the research. L.F. wrote the manuscript with input from all authors.

\bibliography{references}

\clearpage
\onecolumngrid                       

\setcounter{section}{0}\renewcommand{\thesection}{S\arabic{section}}
\setcounter{figure}{0}\renewcommand{\thefigure}{S\arabic{figure}}
\setcounter{table}{0}\renewcommand{\thetable}{S\arabic{table}}
\setcounter{equation}{0}\renewcommand{\theequation}{S\arabic{equation}}

\begin{center}
  {\large\bfseries Supplementary Information:\\[2pt]
  General Noise Channels and Mid-Circuit Measurements for\\
  Pre-Trajectory Sampling with Batched Execution}\\[10pt]
  Louis Follet\textsuperscript{1,2}, Paavai Pari\textsuperscript{1}, Isidor Schoch\textsuperscript{1,3}, Thien Nguyen\textsuperscript{1}, Justin G. Lietz\textsuperscript{1},\\
  Daniel Lowell\textsuperscript{1}, Brucek Khailany\textsuperscript{1}, Krysta M. Svore\textsuperscript{1}, Dirk R. Englund\textsuperscript{2}, and Taylor L. Patti\textsuperscript{1}\\[4pt]
  {\small\textsuperscript{1}NVIDIA Corporation, Santa Clara, CA, USA\\
  \textsuperscript{2}Research Laboratory of Electronics, Massachusetts Institute of Technology, Cambridge, MA 02139, USA\\
  \textsuperscript{3}Department of Materials, University of Oxford, Parks Road, Oxford OX1 3PH, UK}

\end{center}
\vspace{6pt}

\makeatletter
\let\addtocontents\si@origaddtocontents
\makeatother
{\small\setcounter{tocdepth}{2}\tableofcontents}

\section{Exactness of the method}
\label{app:exactness}

\subsection{Events, proposals, and trajectories}
\label{app:event-scope}

The circuit starts in a normalized pure state $|\psi_0\rangle$ and contains $L$ stochastic events, each a local Markovian channel or a measurement. Event $V_v$ has outcomes $a\in\mathcal A_v$ with square Kraus operators $K_{v,a}$, and we write $A_{v,a}=K_{v,a}^\dagger K_{v,a}$, so that $\sum_aA_{v,a}=I$.

A proposal $\boldsymbol Z=(Z_1,\ldots,Z_L)$ carries one variable $Z_v$ per event, either a pre-sampled outcome or the residual distribution $\boldsymbol p^{\mathrm{res}}_v$. A trajectory $\alpha$ lists every outcome needed to reconstruct its state, including reset outcomes that are never recorded. A shot consists of the recorded mid-circuit outcomes and one bitstring $y$ sampled from the trajectory’s final state. Each final state is sampled $m_\alpha$ times, so several shots can share a trajectory.

\subsection{Pre-sampled and deferred outcomes}
\label{app:event-exactness}

Eqs.~\eqref{eq:bounds}--\eqref{eq:decomposition} split each outcome probability into a state-independent part $\pb_{v,a}=\sigma_{\min}(K_{v,a})^2=\lambda_{\min}(A_{v,a})$, the lower bound of the delayed inner product algorithm~\cite{isakov2021qsim}, and a residual part. Two facts complete the argument. First, $A_{v,a}\succeq\pb_{v,a}I$, so for $s_v<1$ the operators
\begin{equation}
 R_{v,a}=\frac{A_{v,a}-\pb_{v,a}I}{1-s_v}
 \label{eq:app-residual-povm}
\end{equation}
are positive semidefinite and sum to the identity. The residual probabilities $p^{\mathrm{res}}_{v,a}(\psi)=\langle\psi|R_{v,a}|\psi\rangle$ of Eq.~\eqref{eq:residual} therefore form a distribution on every normalized state. Second, the $Z_v$ are drawn before execution, independently across events, with
\begin{equation}
 \Pr(Z_v=a)=\pb_{v,a},\qquad \Pr\bigl(Z_v=\boldsymbol p^{\mathrm{res}}_v\bigr)=1-s_v,\qquad P(\boldsymbol Z)=\prod_{v=1}^{L}\Pr(Z_v),
 \label{eq:app-token-law}
\end{equation}
and a deferred event is completed at run time by an independent draw from $\boldsymbol p^{\mathrm{res}}_v$ on the intermediate state.

Both routes apply the update of Eq.~\eqref{eq:physical-state-update}, and by Eq.~\eqref{eq:decomposition} their probabilities add up to $p_{v,a}(\psi)$. Each event therefore selects its outcome and its conditional state with the ordinary trajectory law. Since the variables $Z_v$ are independent, induction over the events extends this to the joint law of all mid-circuit and terminal measurements. The argument holds in exact arithmetic, and Supplementary Note~\ref{app:precision} gives the numerical tolerances.

Any smaller bounds $0\le b_{v,a}\le\pb_{v,a}$ give the same result once $s_v$ and $R_{v,a}$ are recomputed from them. Supplementary Note~\ref{app:conditional-twirl} uses this freedom. At $s_v=0$ every event is deferred and $p^{\mathrm{res}}_{v,a}=p_{v,a}$. At $s_v=1$, completeness forces $A_{v,a}=\pb_{v,a}I$, so the outcome probabilities do not depend on the state and no residual draw occurs.

\subsection{Grouped execution}
\label{app:grouped-exactness}

The $N$ proposals are independent draws, and grouping identical ones extends the trajectory reuse of the original PTSBE~\cite{patti2025ptsbe} algorithm. The $c_g$ members of a group share their outcomes so far, their mid-circuit record, and their remaining proposal, hence the same state $|\psi_g\rangle$. At a deferred event, drawing their outcomes independently would give outcome counts $C_{g,a}$, with $\sum_aC_{g,a}=c_g$, distributed as
\begin{equation}
 \Pr(\{C_{g,a}\}_a\mid c_g,\psi_g)
 =\frac{c_g!}{\prod_a C_{g,a}!}
   \prod_a\left[p^{\mathrm{res}}_{v,a}(\psi_g)\right]^{C_{g,a}}.
 \label{eq:app-multinomial}
\end{equation}
One draw of this vector reproduces these counts, and only outcomes with $C_{g,a}>0$ are evolved. Repeating the argument at every later event shows that grouped and independent execution produce the same law. Two proposals can end in the same trajectory $\alpha$, because an outcome can be pre-sampled or drawn on the intermediate state. They then share a final state and are merged by adding their multiplicities.

The multiplicities sum to $N$, so drawing $c_\alpha$ terminal samples from each final state reproduces the counts of $N$ independent trajectory simulations. Supplementary Note~\ref{app:sampling} treats larger terminal allocations, and Algorithm~\ref{alg:ptsbe} summarizes the procedure.

\subsection{Measurements, resets, and readout}
\label{app:measurement-readout}

A projective measurement with more than one possible outcome has $\lambda_{\min}(P_a)=0$ for every projector, hence $s_v=0$, and its residual law is the Born rule. A reset to $|0\rangle$ uses $K_0=|0\rangle\langle0|$ and $K_1=|0\rangle\langle1|$. Its outcome is not recorded but stays in the trajectory, because for an entangled input the two outcomes leave the other qubits in different states. Terminal measurements of commuting Pauli products are sampled jointly after one basis change, which preserves their correlations.

\section{Kraus representation and \texorpdfstring{max-$s$}{max-s}}
\label{app:maxs}

\subsection{Unitary freedom and objective}
\label{app:maxs-objective}

Suppressing the event index, let a channel $\mathcal E$ on a $d$-dimensional space have Kraus operators $K_1,\ldots,K_k$. Any unitary $W\in\mathrm U(k)$ gives another representation $K'_a=\sum_bW_{ab}K_b$ of the same channel, since
\begin{equation}
 \sum_aK'_a\rho K_a^{\prime\dagger}
 =\sum_{b,c}\Bigl(\sum_aW_{ab}W_{ac}^*\Bigr)K_b\rho K_c^\dagger
 =\sum_bK_b\rho K_b^\dagger .
 \label{eq:app-remix-invariance}
\end{equation}
The trajectories change with the representation~\cite{vovk2022entanglement, chen2024unraveling, daraban2025numu, cichy2025locally}, and so does the pre-sampling probability. Max-$s$ preprocessing searches for the representation with the largest
\begin{equation}
 s(W)=\sum_a\sigma_{\min}\Bigl(\sum_bW_{ab}K_b\Bigr)^2.
 \label{eq:app-maxs}
\end{equation}
It applies only to noise channels whose outcomes are not recorded, since remixing the outcomes of a measurement would change the record. A larger $s_v$ leaves fewer deferred draws, $\sum_v(1-s_v)$ per proposal on average.

\subsection{Upper bound and amplitude damping}
\label{app:maxs-bound}

The matrix $K_aK_a^\dagger$ has the same eigenvalues as $K_a^\dagger K_a$, so $K_aK_a^\dagger\succeq\pb_aI$. Summing over $a$ gives $\mathcal E(I)\succeq sI$, hence
\begin{equation}
 s\le s_{\mathrm{ub}}\equiv\lambda_{\min}\!\left[\mathcal E(I)\right]\le1
 \label{eq:app-s-bound}
\end{equation}
for every representation, where the last inequality follows from $\operatorname{Tr}\mathcal E(I)=d$. A representation that reaches $s_{\mathrm{ub}}$ is optimal.

If $s=1$, the positive semidefinite matrices $K_a^\dagger K_a-\pb_aI$ sum to zero and all vanish. Each nonzero $K_a$ is then $\sqrt{\pb_a}$ times a unitary, and the representation is a mixture of unitaries. Conversely, a channel that is a mixture of unitaries has $s=1$ in its mixture form, even when another of its representations has $s<1$. Nonunital channels never reach $s=1$, since a mixture of unitaries maps $I$ to $I$.

The standard amplitude-damping representation of Eq.~\eqref{eq:app-ad-kraus} has $\pb_0=1-\gamma$ and $\pb_1=0$, and
\begin{equation}
 \mathcal E(I)=\operatorname{diag}(1+\gamma,\,1-\gamma),
 \qquad s=s_{\mathrm{ub}}=1-\gamma .
 \label{eq:app-ad-optimal}
\end{equation}
It is therefore optimal, and preprocessing leaves it unchanged. The Pauli channels of the random circuits already have $s=1$, so preprocessing plays no role in the reported timings. For generalized amplitude damping with equilibrium excited population $p_e<1/2$ (Sec.~\ref{app:maxs-examples}), $\mathcal E(I)=\operatorname{diag}[1+\gamma(1-2p_e),\,1-\gamma(1-2p_e)]$ and $s_{\mathrm{ub}}=1-\gamma(1-2p_e)$.

\subsection{Numerical search}
\label{app:maxs-search}

The search starts from the supplied representation and from random unitaries. Each step multiplies $W$ by $e^{i\varepsilon H}$ for a random Hermitian $H$, keeps the step only if $s$ increases, and shrinks $\varepsilon$ after a rejection. It stops within $10^{-9}$ of the bound~\eqref{eq:app-s-bound} or after a fixed number of steps. A candidate must reproduce the Choi matrix $\sum_a\operatorname{vec}(K_a)\operatorname{vec}(K_a)^\dagger$ of the input to $10^{-10}$ and pass the checks of Supplementary Note~\ref{app:precision}. Otherwise the original representation is kept, so the adopted $s$ never decreases.

\subsection{Examples}
\label{app:maxs-examples}

Table~\ref{tab:app-maxs} applies the search to five channels.
\begin{itemize}
\item Amplitude damping, Eq.~\eqref{eq:app-ad-kraus}, is already optimal (Sec.~\ref{app:maxs-bound}).
\item Phase damping with parameter $\lambda$, with $K_0=\operatorname{diag}(1,\sqrt{1-\lambda})$ and $K_1=\operatorname{diag}(0,\sqrt\lambda)$, multiplies the coherences by $\sqrt{1-\lambda}$. So does the mixture $K'_0=\sqrt{1-q}\,I$, $K'_1=\sqrt q\,Z$ with $q=(1-\sqrt{1-\lambda})/2$, which has $s=1$.
\item Generalized amplitude damping with equilibrium excited population $p_e$ uses $\sqrt{1-p_e}\,K_0$, $\sqrt{1-p_e}\,K_1$, $\sqrt{p_e}\,XK_0X$, and $\sqrt{p_e}\,XK_1X$, with $K_{0,1}$ the amplitude-damping operators of Eq.~\eqref{eq:app-ad-kraus}.
\item Two-qubit dephasing, with $K_0=\operatorname{diag}(1,1,1,\sqrt{1-\lambda})$ and $K_1=\sqrt\lambda\,|11\rangle\langle11|$, attenuates the coherences involving $|11\rangle$. It has the same closed form as phase damping, with $\operatorname{diag}(1,1,1,-1)$ in place of $Z$.
\item The random rank-three channel draws three $2\times2$ matrices $G_a$ with independent standard normal real and imaginary entries and sets $K_a=G_a\bigl(\sum_bG_b^\dagger G_b\bigr)^{-1/2}$.
\end{itemize}

\begin{table}[tbp]
\centering
\small
\caption{Pre-sampling probability $s$ before and after the search of Sec.~\ref{app:maxs-search}, and the bound $s_{\mathrm{ub}}$ of Eq.~\eqref{eq:app-s-bound}. For phase damping and two-qubit dephasing, the closed forms of Sec.~\ref{app:maxs-examples} reach $s=1$ exactly. Every adopted representation reproduces the input Choi matrix to $10^{-10}$.}
\label{tab:app-maxs}
\begin{tabular}{@{}lccc@{}}
\toprule
Channel & Input & After search & $s_{\mathrm{ub}}$ \\
\midrule
Amplitude damping, $\gamma=0.3$ & 0.700 & 0.70000 & 0.70000 \\
Phase damping, $\lambda=0.4$ & 0.600 & $1-1.3\times10^{-5}$ & 1 \\
Generalized damping, $\gamma=0.2$, $p_e=0.3$ & 0.800 & 0.91985 & 0.92000 \\
Two-qubit dephasing, $\lambda=0.2$ & 0.800 & $1-2.6\times10^{-7}$ & 1 \\
Random channel, rank three & 0.345 & 0.86311 & 0.86313 \\
\bottomrule
\end{tabular}
\end{table}

\section{Terminal allocation}
\label{app:sampling}

\subsection{Proportional mode}
\label{app:proportional-allocation}

Let $P(\alpha)$ be the probability of the trajectory $\alpha$, mid-circuit outcomes included, and $P(y\mid\alpha)$ that of the bitstring $y$ given $\alpha$. The $N$ proposals give $N$ independent trajectory draws with multiplicities $c_\alpha$, and the terminal samples are independent given the trajectories. Supplementary Note~\ref{app:grouped-exactness} shows that $m_\alpha=c_\alpha$ reproduces independent sampling.

The allocation $m_\alpha=bc_\alpha$, with $b$ a positive integer, gives $bN$ records. It corresponds to the optional proportional resampling of the original PTSBE~\cite{patti2025ptsbe}, with sampled multiplicities in place of trajectory probabilities. A record picked at random follows the circuit distribution, but records drawn from the same trajectory are correlated. For a real statistic $f(\alpha,y)$ with mean $\mu_\alpha$ and variance $\sigma_\alpha^2$ given $\alpha$, the average $\widehat\mu$ of $f$ over the $bN$ records has variance
\begin{equation}
 \operatorname{Var}(\widehat\mu)
 =\frac{\operatorname{Var}_{\alpha}(\mu_\alpha)}{N}
 +\frac{\mathbb E_{\alpha}(\sigma_\alpha^2)}{Nb}.
 \label{eq:app-proportional-variance}
\end{equation}
A larger $b$ shrinks the second term. The first term depends only on the number of trajectory draws.

\subsection{Nonproportional mode}
\label{app:flat-allocation}

Nonproportional mode gives the same number $m$ of terminal samples to each of the $N_{\mathrm{traj}}$ distinct trajectories. An unweighted average over the records then weights every trajectory by $1/N_{\mathrm{traj}}$ instead of $c_\alpha/N$. The multiplicities are kept, and with $\overline f_\alpha$ the mean of $f$ over the $m$ samples of $\alpha$, the weighted estimator
\begin{equation}
 \widehat\mu_{\mathrm w}=\sum_{\alpha}\frac{c_\alpha}{N}\,\overline f_\alpha
 \label{eq:app-weighted-estimator}
\end{equation}
is unbiased over the trajectory draws. Its variance has the same trajectory term $\operatorname{Var}_\alpha(\mu_\alpha)/N$ as Eq.~\eqref{eq:app-proportional-variance}, plus a measurement term proportional to $1/m$. When the proposals are chosen deliberately, as in Supplementary Note~\ref{app:stratified-estimation} where each error set $E$ receives its own number of runs, the records of each error set are averaged first and the averages are weighted by $P(E)$, as in Eq.~\eqref{eq:app-stratified-estimator}.

In Fig.~\ref{Fig:Fig4}, $N=N_{\mathrm{traj}}=100$, so all weights are equal.

\subsection{Fraction of distinct shots}
\label{app:distinct-records}

The fraction of distinct shots is the number of distinct pairs $(\alpha,y)$ among the $M=\sum_\alpha m_\alpha$ shots, divided by $M$. Two shots with the same bitstring and different trajectories count as distinct. The fraction measures repetition in the labeled dataset and says little about how often trajectories repeat. At $n=24$ and $D=16$, almost every shot is distinct even though each trajectory serves about four shots (Fig.~\ref{fig:supp-reuse}).

For one trajectory sampled $m$ times, the expected number of distinct bitstrings is
\begin{equation}
 \mathbb E[\nu_\alpha\mid\alpha]
 =\sum_y\left[1-\left(1-P(y\mid\alpha)\right)^m\right].
 \label{eq:app-occupancy}
\end{equation}
A uniform distribution over $2^{30}$ bitstrings would still give a distinct fraction of about 0.995 at $m=10^7$. The measured 0.89 in Fig.~\ref{Fig:Fig4} shows that each trajectory's readout distribution is far from uniform.

\section{Amplitude damping and its Pauli twirl}
\label{app:noise}

\subsection{Memory circuit and noise}
\label{app:qec-circuit}

Fig.~\ref{Fig:Fig2} uses a distance-three rotated surface-code memory storing $\lvert0\rangle_L$, the distance-three $Z$ memory of the Clifft benchmark suite~\cite{clifft} with its noise instructions removed. Nine data qubits and eight ancillas run three rounds of stabilizer measurement, followed by a $Z$-basis measurement of the data qubits. The syndrome $S\in\{0,1\}^{24}$ collects all detectors: four at the first round, eight comparing each pair of successive rounds, and four at the final readout. Each detector, like the logical bit $L$, is a parity of measurement outcomes that vanishes without errors.

Each round lasts $1.1~\mu $s, the cycle time of the distance-three experiment of Ref.~\cite{krinner2022}, and uses that experiment's gate and readout durations: two 40-ns single-qubit layers, four 98-ns entangling layers, and a 400-ns readout interval. The remaining 228 ns form the reset interval. Amplitude damping acts on every qubit after each gate layer and before readout, and on the nine data qubits during reset, which gives $3[(2+4+1)\cdot17+9]=384$ noise locations. An exposure of duration $\Delta t_v$ gives the damping parameter
\begin{equation}
 \gamma_v=1-\exp(-\Delta t_v/T_1),
 \label{eq:app-gamma}
\end{equation}
with $T_1=68\,\mu\mathrm{s}$, the mean operating $T_1$ of the 105-qubit processor of Ref.~\cite{google2025below}. We omit additional dephasing, so the model has $T_2=2T_1$, and amplitude damping is the only noise source. Gates, initialization, and measurements are otherwise ideal, and the exact and twirled simulations share the circuit, the detectors, and the noise locations.

\subsection{Twirl formulas}
\label{app:pauli-twirl}

At location $v$, amplitude damping has Kraus operators
\begin{equation}
 K_{v,0}=\operatorname{diag}\bigl(1,\sqrt{1-\gamma_v}\bigr),\qquad
 K_{v,1}=\sqrt{\gamma_v}\,|0\rangle\langle1| .
 \label{eq:app-ad-kraus}
\end{equation}
A qubit with excited population $\rho_{11}$ decays with probability $\gamma_v\rho_{11}$. The channel leaves $\lvert0\rangle$ unchanged and is nonunital, since it maps $I$ to $I+\gamma_vZ$. The Pauli twirl of a single-qubit channel~\cite{gellerzhou2013}, $\mathcal T(\mathcal E)(\rho)=\tfrac14\sum_PP\,\mathcal E(P\rho P)\,P$ with $P\in\{I,X,Y,Z\}$, removes the cross terms between the Pauli components of the Kraus operators. It leaves the Pauli channel with probabilities $p_P=\sum_a|\operatorname{Tr}(PK_a)/2|^2$, which for amplitude damping are
\begin{equation}
 p_X=p_Y=\frac{\gamma_v}{4},\qquad
 p_Z=\frac{\bigl(1-\sqrt{1-\gamma_v}\bigr)^2}{4},\qquad
 p_I=1-\frac{\gamma_v}{2}-p_Z .
 \label{eq:app-ad-twirl}
\end{equation}
The twirled channel is unital and flips $\lvert0\rangle$ with probability $\gamma_v/2$, where damping does nothing. Using Eq.~\eqref{eq:app-ad-twirl} at every location gives the Pauli circuit that we sample with Stim~\cite{gidney2021stim}.

\subsection{Conditioning on an error set}
\label{app:conditional-twirl}

The standard representation has $\pb_{v,0}=1-\gamma_v$ and $\pb_{v,1}=0$, so each location is deferred with probability $\gamma_v$. The error set $E$ of a proposal is its set of deferred locations, which need not coincide with the locations where a decay occurs. It has probability
\begin{equation}
 P(E)=\prod_{v\in E}\gamma_v\prod_{v\notin E}(1-\gamma_v).
 \label{eq:app-error-set-probability}
\end{equation}
Outside $E$ the update $K_{v,0}$ is applied, and inside $E$ the qubit decays with probability $\rho_{11}$ on the intermediate state.

Fig.~\ref{Fig:Fig2}(b) conditions the twirled channel on the same $E$ by applying Supplementary Note~\ref{app:event-exactness} to the twirled channel with bounds $b_I=1-\gamma_v$ and $b_X=b_Y=b_Z=0$. The twirled model then applies the identity outside $E$ and, inside $E$, a Pauli drawn from the residual distribution
\begin{equation}
 (\pi_I,\pi_X,\pi_Y,\pi_Z)
 =\left(\frac12-\frac{p_Z}{\gamma_v},\ \frac14,\ \frac14,\ \frac{p_Z}{\gamma_v}\right).
 \label{eq:app-conditional-twirl}
\end{equation}
Supplementary Note~\ref{app:exactness} then gives
\begin{equation}
 \sum_E P(E)P_{\mathrm{twirl}}(S\mid E)=P_{\mathrm{twirl}}(S),
 \label{eq:app-twirl-mixture}
\end{equation}
with the same $P(E)$ as the exact model. Other bounds give the same unconditional twirl with different conditional distributions, so the per-set comparison depends on this choice, while the global comparisons do not.

Each injected Pauli flips a fixed set of detectors, so summing over the at most $4^{|E|}$ assignments gives $P_{\mathrm{twirl}}(S\mid E)$ exactly. Under exact noise the $K_{v,0}$ updates outside $E$ still act, so even the empty set gives a detection, with probability $6.3\times10^{-4}$, whereas its twirled distribution is always quiet.

\section{Statistics of the QEC comparison}
\label{app:statistics}

\subsection{Estimate of the exact distribution}
\label{app:stratified-estimation}

Our method generates the exact-noise records in single precision, with one terminal sample per draw, each record being a pair $(S,L)$. We estimate their distribution error set by error set. Every set with $|E|\le2$ is simulated on its own, and sets with $|E|\ge3$ are drawn from Eq.~\eqref{eq:app-error-set-probability} conditioned on $|E|\ge3$ and pooled into one tail sample (Table~\ref{tab:app-strata}). Counts are kept separately for each error set, even when two error sets lead to the same trajectory. With $\widehat q_E(x)$ the fraction of the $n_E$ records of error set $E$ equal to $x$, and $\widehat q_{\mathrm{tail}}(x)$ the same fraction among the $n_{\mathrm{tail}}$ records with $|E|\ge3$,
\begin{equation}
 \widehat P_{\mathrm{exact}}(x)
 =\sum_{|E|\le2}P(E)\,\widehat q_E(x)
  +P(|E|\ge3)\,\widehat q_{\mathrm{tail}}(x)
 \label{eq:app-stratified-estimator}
\end{equation}
for any value $x$ of $S$ or of $(S,L)$. The weights are known exactly, so the allocation sets the precision without biasing the estimate. Sampling is independent within and across error sets, which gives the plug-in variance
\begin{equation}
 \widehat{\operatorname{Var}}(x)
 =\sum_{|E|\le2}\frac{P(E)^2}{n_E}\,\widehat q_E(x)[1-\widehat q_E(x)]
 +\frac{P(|E|\ge3)^2}{n_{\mathrm{tail}}}\,\widehat q_{\mathrm{tail}}(x)[1-\widehat q_{\mathrm{tail}}(x)].
 \label{eq:app-stratified-variance}
\end{equation}

\begin{table}[tb]
 \centering
 \small
 \caption{Exact-noise allocation for Fig.~\ref{Fig:Fig2}. The first three rows cover every set of their size, 96.1\% of the probability. The global distribution also uses $3.5\times10^6$ records with $|E|\ge3$. The last row is the separate sample for panel (b), in which 4,000 draws gave 3,995 distinct sets.}
 \label{tab:app-strata}
 \begin{tabular}{crrr}
 \toprule
 $|E|$ & Sets simulated & Records per set & $P(|E|)$\\
 \midrule
 0 & 1 & 17,893,376 & 0.475\\
 1 & 384 & 1,116,160 & 0.354\\
 2 & 73,536 & 2,048 & 0.131\\
 3 & 3,995 & 2,048 & 0.032\\
 \bottomrule
 \end{tabular}
\end{table}

Fig.~\ref{Fig:Fig2}(a) uses $10^9$ Stim records and Fig.~2(c) uses $2.1\times10^{10}$. A separate set of $10^7$ exact-noise records from proportional sampling serves as the decoder test set and checks this estimate: the two quiet-syndrome probabilities differ by 0.3 combined standard errors. Rebuilding the twirled distribution from its conditional distributions through Eq.~\eqref{eq:app-twirl-mixture} also agrees with direct Stim sampling, with standardized differences of mean 0.02 and standard deviation 0.97 over the 1,000 most likely nonzero syndromes.

\subsection{Syndrome ratios and distances}
\label{app:syndrome-statistics}

Fig.~\ref{Fig:Fig2}(a) ranks the nonzero syndromes by $\tfrac12[\widehat P_{\mathrm{exact}}(S)+\widehat P_{\mathrm{twirl}}(S)]$ and shows the first 1,000, which carry 95\% of the exact and 93\% of the twirled probability given a detection. It plots the ratio $\widehat P_{\mathrm{twirl}}(S)/\widehat P_{\mathrm{exact}}(S)$, with pointwise bars $e^{\pm\widehat\sigma_{\log}}$ around it, where
\begin{equation}
 \widehat\sigma_{\log}^2(S)
 =\frac{\widehat{\operatorname{Var}}(S)}{\widehat P_{\mathrm{exact}}(S)^2}
  +\frac{1-\widehat P_{\mathrm{twirl}}(S)}{c_{\mathrm{twirl}}(S)}
\end{equation}
and $c_{\mathrm{twirl}}(S)$ counts $S$ in the twirled sample. Among the displayed syndromes seen under both models, the largest ratio is 329, with $\widehat\sigma_{\log}=0.21$, and 11 ratios exceed 100. Another 25 displayed syndromes never occur in the exact sample and are marked separately.

A detection occurs in 25.1\% of the exact shots and 30.3\% of the twirled ones. We compute the total-variation distance $d_{\mathrm{TV}}(P,Q)=\tfrac12\sum_S|P(S)-Q(S)|$ between the full distributions and between the distributions given a detection (Table~\ref{tab:app-tvd}). For the uncertainty, 200 Poisson bootstrap replicates replace every count by a Poisson draw with that mean and renormalize each error set and the tail sample with their weights fixed. The standard error is the standard deviation of the replicate distances. Their mean exceeds the observed distance by 0.02 percentage points over all syndromes and 0.05 given a detection, which is how much finite sampling inflates the estimate. The noise reference is the 95th percentile over replicates of $\tfrac12\sum_S|\delta_{\mathrm{exact}}(S)-\delta_{\mathrm{twirl}}(S)|$, where $\delta$ is the resampled minus the original estimate. It gives the distance that sampling alone produces.

\begin{table}[tb]
 \centering
 \small
 \caption{Total-variation distances between the exact and twirled syndrome distributions of Fig.~\ref{Fig:Fig2}, in percent.}
 \label{tab:app-tvd}
 \begin{tabular}{lrr}
 \toprule
 & All syndromes & Given a detection\\
 \midrule
 Estimate & 6.54 & 19.84\\
 Bootstrap standard error & 0.001 & 0.004\\
 Noise reference & 0.10 & 0.38\\
 \bottomrule
 \end{tabular}
\end{table}

\subsection{Distances given the error set}
\label{app:conditional-distances}

Fig.~\ref{Fig:Fig2}(b) shows, for every set with at most two deferred locations and for the three-location sample, the distance $d_E=d_{\mathrm{TV}}\bigl(\widehat P_{\mathrm{exact}}(\cdot\mid E),P_{\mathrm{twirl}}(\cdot\mid E)\bigr)$, with tied ranks spread at random. Weighted by $P(E)$, or by draw multiplicity in the three-location sample, its mean is 0.06\%, 9.9\%, 20.1\%, and 28.8\% for $|E|=0$, 1, 2, and 3. If the exact conditional distributions equaled the twirled ones, finite sampling alone would give 0, 0.04\%, 1.9\%, and 3.1\%, the same weighted mean of the median distance between each set's twirled distribution and 200 multinomial samples of the exact sample size. These references are not subtracted. The three examples are the most probable sets whose distance lies within 0.02 of 0.5, 0.75, and 0.25.

\subsection{Logical flips and decoders}
\label{app:decoder-evaluation}

Fig.~\ref{Fig:Fig2}(c) plots each model's estimate of $P(L=1\mid S)=\widehat P(S,L=1)/\widehat P(S)$ for the syndromes with at least 100 twirled records and at least 100 effective exact records, $\widehat P_{\mathrm{exact}}(S)^2/\widehat{\operatorname{Var}}(S)$, among the quiet syndrome and the $10^4$ most likely nonzero ones. The 4,588 syndromes kept carry 99.75\% of the exact probability. On the twirled side, 100 records bound the binomial standard error of $P(L=1\mid S)$ by 0.05. On the exact side, the delta-method standard error is at most 0.05 for every displayed syndrome.

Each lookup decoder applies the logical correction when its model gives $P(L=1\mid S)>1/2$, and no correction for ties or for syndromes absent from the data used to build it. We freeze both tables and evaluate them on the same $10^7$ independent exact-noise records. The table built from exact statistics fails on $(0.424\pm0.002)\%$ of them and the table built from twirled statistics on $(0.458\pm0.002)\%$. The paired difference, $0.034\pm0.001$ percentage points, is a relative increase of 7.9\%. The uncertainties are one standard error over the test records, with the tables held fixed, and do not include the finite-sample error of the tables themselves, which could flip syndromes whose estimates of $P(L=1\mid S)$ lie close to $1/2$. The 53 circled syndromes, whose estimates of $P(L=1\mid S)$ lie on opposite sides of $1/2$, carry 0.11\% of the exact probability.

\section{Implementation}
\label{app:implementation}

\subsection{Execution and caching}
\label{app:cache}

Three random-number streams derived from one seed serve the proposals, the residual draws, and the terminal samples, so changing the terminal allocation leaves the trajectories unchanged. Groups are traversed depth first. At a branch, the first occupied outcome continues from the intermediate state, and each other one keeps its outcomes $Z_1$, …, $Z_v$ and a reference to a copy of the state before the branch. In the benchmarks these copies stay on the GPU within a memory budget, and the oldest are evicted when memory runs out. A missing copy is rebuilt by re-applying $Z_1$, …, $Z_{v-1}$ without new draws, so the cache capacity changes the run time and leaves the results unchanged.

The number of distinct proposals limits the reuse. For $N$ independent draws, its expectation is
\begin{equation}
 \mathbb E[N_{\mathrm{prop}}]
 =\sum_{\boldsymbol Z}\left[1-\left(1-P(\boldsymbol Z)\right)^N\right],
 \label{eq:app-proposal-occupancy}
\end{equation}
much smaller than $N$ when a few proposals dominate, while branching at deferred events raises the number of distinct trajectories again. Figure~\ref{fig:supp-reuse} shows the result for the random circuits. Doubling the depth gives about ten times more distinct trajectories per shot. A deferred damping location that ends without a decay applies $K_0$, as a pre-sampled one does, so proposals that differ only at such locations merge into one trajectory. The amplitude-damping circuits accordingly give two to four times fewer distinct trajectories than their Pauli counterparts.

\subsection{Finite precision}
\label{app:precision}

Channel preprocessing runs in double precision and the statevectors in single precision. The state is not renormalized after every operator. If the stored state is $|\phi\rangle=\kappa|\psi\rangle$, then
\begin{equation}
 \frac{\langle\phi|R_{v,a}|\phi\rangle}
      {\sum_b\langle\phi|R_{v,b}|\phi\rangle}
 =\langle\psi|R_{v,a}|\psi\rangle,
 \label{eq:app-norm-ratio}
\end{equation}
because the residual operators sum to the identity, so residual probabilities are read without renormalizing. Pre-sampled outcomes apply $K_{v,a}/\sqrt{\pb_{v,a}}$, which limits the loss of norm along long sequences of pre-sampled outcomes and leaves the normalized state unchanged.

Each channel must preserve the trace to $10^{-9}$, and when $1-s_v>10^{-9}$ its residual operators must sum to the identity to $10^{-7}$ in spectral norm. Below $1-s_v=10^{-9}$ the residual branch is dropped. In exact arithmetic, this changes the outcome distribution of that event by at most $1-s_v$ in total variation, and that of the circuit by at most the sum of the dropped masses. For small circuits, host-side tests compare the probability of every measurement record, summed over the pre-sampled and deferred routes, with a density-matrix calculation. They also check trace preservation, the bound decomposition, and the invariance of the Choi matrix under remixing.

\section{Benchmarks}
\label{app:benchmarks}

\subsection{Circuits and noise}
\label{app:benchmark-circuits}

The memory benchmarks use the surface-code $Z$ memory of Supplementary Note~\ref{app:qec-circuit}, the $X$ memory with the same layout and damping schedule and with $X$-basis preparation and readout, and a distance-three color-code memory generated with Stim~\cite{gidney2021stim}. The color code has seven data qubits and three ancillas. Each round applies $S^\dagger$ and $H$ to the data qubits, then six CNOT layers and ancilla measurement and reset, with intervals of 40~ns, $6\times98$~ns, 400~ns, and 228~ns. Damping acts on all ten qubits during the first eight intervals and on the data qubits during reset, which gives $3(8\times10+7)=261$ locations. All three memories use $T_1=68\,\mu\mathrm{s}$ and run three rounds.

\begin{figure*}[tbp]
 \centering
 \includegraphics[width=\textwidth]{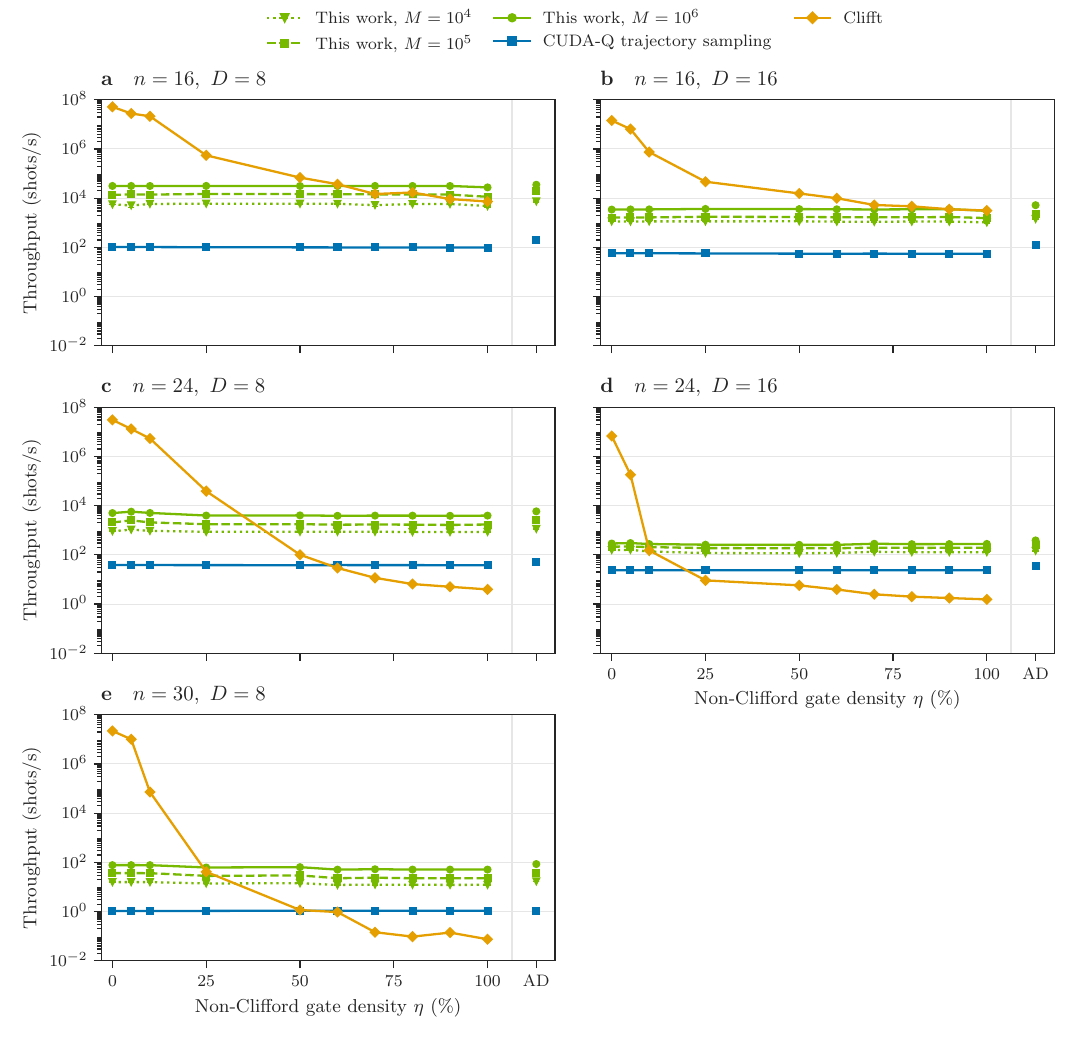}
 \caption{\textbf{Throughput for the random monitored circuits.}
 Green curves are our method in proportional mode at $M=10^4$, $10^5$, and $10^6$ shots, blue curves CUDA-Q, and orange curves Clifft with 16 workers. AD denotes the $\eta=1$ circuits with amplitude damping in place of the Pauli gate noise. Points are geometric means over three runs, and bars span one standard deviation of the log rate.}
 \label{fig:supp-throughput-grid}
\end{figure*}

\begin{figure*}[tbp]
 \centering
 \includegraphics[width=\textwidth]{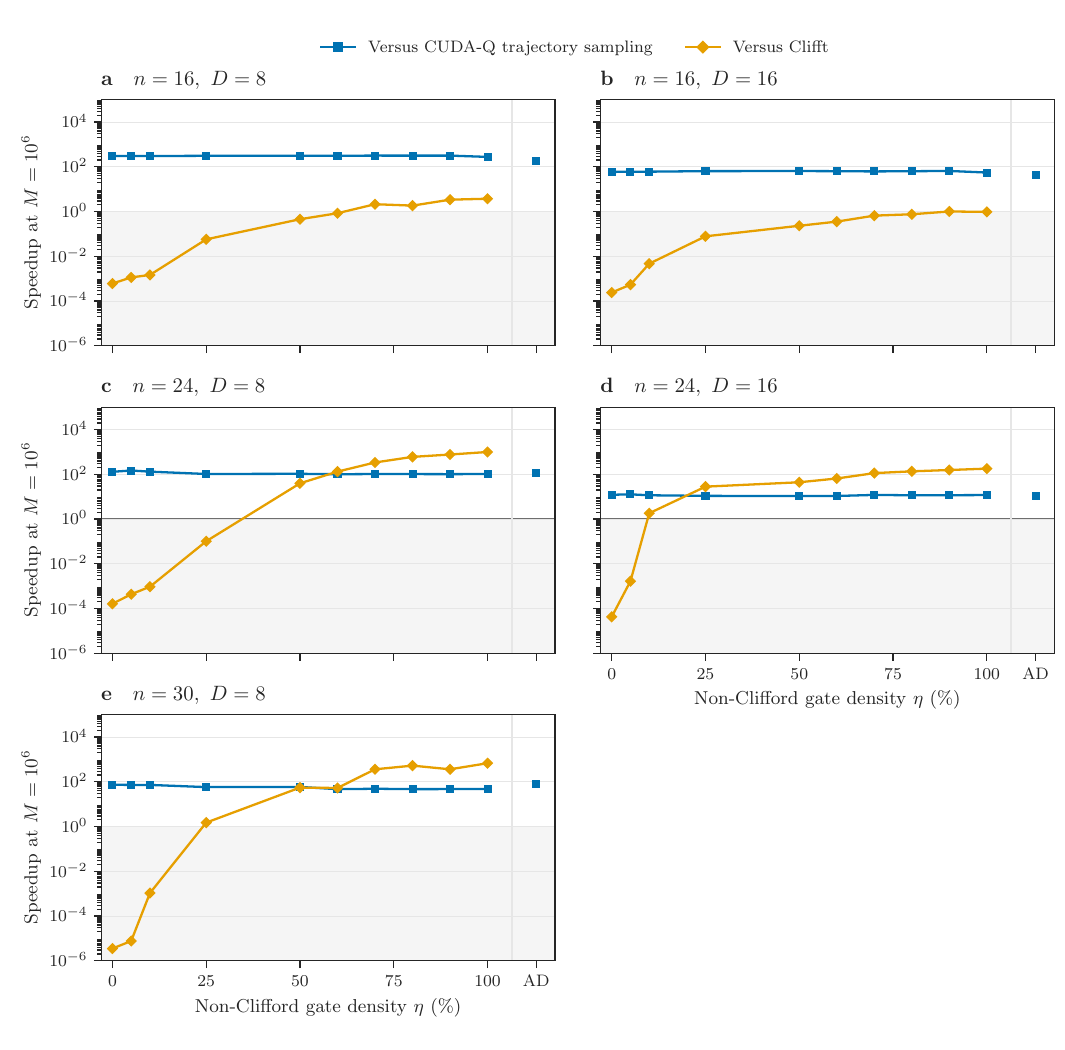}
 \caption{\textbf{Speedup at $10^6$ proportional shots.}
 Ratios of the rates of Fig.~\ref{fig:supp-throughput-grid} to those of CUDA-Q (blue) and Clifft (orange), with error bars adding the two log standard deviations in quadrature. Our method is faster in the white region.}
 \label{fig:supp-speedup-grid}
\end{figure*}

\begin{figure*}[tbp]
 \centering
 \includegraphics[width=\textwidth]{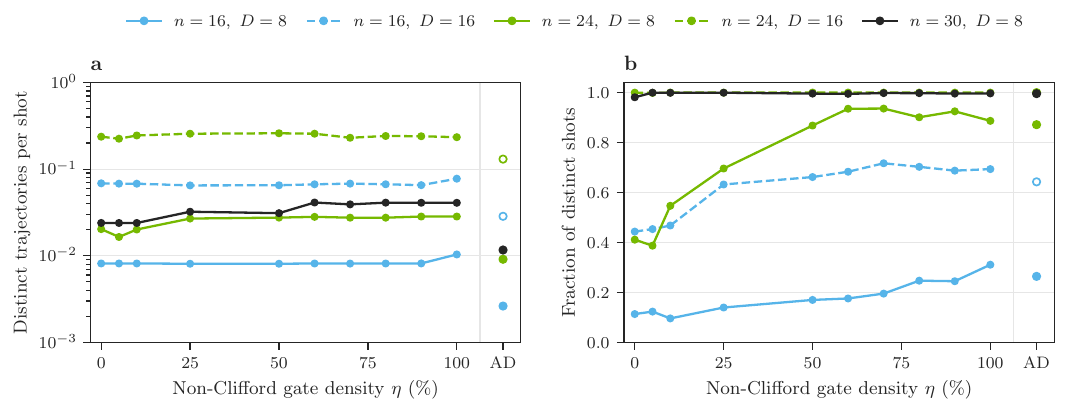}
 \caption{\textbf{Trajectory reuse in the random monitored circuits at $10^6$ proportional shots.}
 \textbf{a,} Distinct trajectories per shot. Each distinct trajectory is prepared once for terminal sampling. \textbf{b,} Fraction of distinct shots, the number of distinct (trajectory, measurement outcome) pairs per shot. At $n=24$ and $D=16$ almost every shot is distinct, even though each trajectory serves about four shots. Open markers are the AD circuits at $D=16$. Means over three runs.}
 \label{fig:supp-reuse}
\end{figure*}

The random monitored circuits have $n\in\{16,24,30\}$ qubits and $D\in\{8,16\}$ alternating layers of single-qubit gates from $\{H,S,S^\dagger\}$ and two-qubit gates CX or CZ on random pairs. Each gate is replaced, with probability $\eta$, by a rotation $R_P(\theta)=\exp(-i\pi\theta P/2)$ about an axis $P$ drawn from $\{X,Y,Z\}$ or $\{XX,YY,ZZ\}$. The angle $\theta$ is drawn uniformly from $[0.08,0.92]$, kept at least 0.05 away from multiples of 0.5, and rounded to three decimals. One random instance is fixed for each $n$ and $D$, and the rotated gates are nested as $\eta$ grows over $\{0,0.05,0.1,0.25,0.5,0.6,0.7,0.8,0.9,1\}$, so raising $\eta$ only replaces more gates. After every second layer except the last, $\lceil0.05n\rceil$ qubits are measured and reset, and all qubits are measured at the end, which records $n+(D/2-1)\lceil0.05n\rceil$ bits per shot. Main-text Fig.~\ref{Fig:Fig3}(b) shows $D=8$, and Figs.~\ref{fig:supp-throughput-grid}--\ref{fig:supp-reuse} add $D=16$ for $n=16$ and 24, where deeper circuits give less reuse and smaller speedups.

The noise is the same for every $\eta$. Each single-qubit layer is followed by a Pauli channel of strength $10^{-3}$ on every qubit, cycling through depolarizing, $X$, $Y$, and $Z$ errors across the gate locations. Each two-qubit layer is followed by two-qubit depolarizing noise of strength $10^{-3}$ on every pair, and a bit flip of probability $10^{-3}$ precedes every measurement. A Pauli error of strength $p$ applies its Pauli with probability $p$, and depolarizing noise of strength $p$ applies each nonidentity Pauli with probability $p/3$ on one qubit and $p/15$ on two. The AD circuits keep the $\eta=1$ gates and the bit flips and replace the gate noise by amplitude damping, with $\gamma=10^{-3}$ after single-qubit layers and $5\times10^{-4}$ per qubit after two-qubit layers. All engines read the same circuit files.

\subsection{Engines and timing}
\label{app:benchmark-timing}

The reported runs used an earlier version of our implementation, which drives an internal NVIDIA GPU statevector backend directly. The released code will use the public cuStateVec library instead, with comparable timings in our tests. The reference calls \texttt{cudaq.run} with the \texttt{nvidia} target of CUDA-Q~\cite{cudaq}, which returns the full measurement record of every shot, mid-circuit outcomes included. It launches the kernel once per shot, so CUDA-Q evolves one trajectory per shot. Both engines use single-precision statevectors and simulate the exact channels, so the speedups combine trajectory reuse with the difference between the two execution paths.

Our execution time is $t_{\mathrm{exec}}=t_{\mathrm{generation}}+t_{\mathrm{sample}}$. Trajectory generation covers proposal sampling, grouping, grouped evolution, and branching. Sampling covers the re-execution of each trajectory to its final state and the terminal draws. A future version of our implementation will sample each trajectory as soon as its evolution ends, which avoids this re-execution. Circuit parsing, channel preparation including max-$s$ preprocessing, backend initialization, and the formatting and writing of the output are left out of the timer. All proportional benchmarks use $m_\alpha=c_\alpha$, so a run of $N$ draws produces $M=N$ shots at a rate $M/t_{\mathrm{exec}}$.

The CUDA-Q reference warms up, sets a batch size from a short probe, and times fixed-size batches over a 120-s window, or a 30-min window at $n=30$, on one GPU. Its rate is the mean over batches of shots per batch time, without compilation and loading. The Clifft reference~\cite{clifft} runs 16 single-threaded workers pinned to distinct cores, each with its own compiled circuit, over 120-s windows after an untimed probe, and its rate is the sum of the workers' mean rates. Clifft does not support amplitude damping.

\subsection{Hardware and cache use}
\label{app:benchmark-hardware}

The proportional benchmarks ran on exclusive nodes with eight NVIDIA H100 80-GB GPUs and two Intel Xeon Platinum 8480CL CPUs, with one GPU per run. Bulk generation used one H100 GPU of a node of the same type.  
State copies are kept on the GPU, and before a new copy is stored, the oldest are evicted until at least $2\times10^9$ bytes would remain free (Supplementary Note~\ref{app:cache}). A copy takes $8\cdot2^n$ bytes, 8~GiB at $n=30$. The copies held at any time stayed below 1.7~GiB at $n\le24$, reached 16 to 48~GiB at $n=30$, and reached 56~GiB, seven copies, in bulk generation. No run evicted a copy.

\subsection{Run conditions and error bars}
\label{app:benchmark-statistics}

GPU clocks and thermal state were logged and competing GPU use was monitored in every session. The Clifft windows ran one at a time.

Each proportional point has three runs. The random-circuit runs of both GPU engines use three different sampling seeds. The QEC runs of our method repeat one seed, which measures timing variation on identical data, and their CUDA-Q references use three seeds. The Clifft windows are timing repetitions. Points are geometric means of the three rates, with bars spanning one sample standard deviation $\sigma_{\log}$ of the log rate on each side. A speedup is the ratio of two geometric means, with $\sigma_{\log}=(\sigma_{\log,\mathrm{ours}}^2+\sigma_{\log,\mathrm{ref}}^2)^{1/2}$, which assumes that the engines vary independently. The bars show the run-to-run spread on fixed circuits.

\subsection{Bulk generation}
\label{app:benchmark-bulk}

The bulk-generation circuit has $n=30$, $D=16$, $\eta=1$, and amplitude damping, and records 14 mid-circuit and 30 terminal bits per shot. The 100 proposals give $N_{\mathrm{traj}}=100$ distinct trajectories, each sampled $m$ times. Each budget repeats one seed three times, and the CUDA-Q reference uses three seeds over 30-min windows. The plotted rate per trajectory is $m/t_{\mathrm{exec}}$, the total shot rate $N_{\mathrm{traj}}m/t_{\mathrm{exec}}$ divided by $N_{\mathrm{traj}}$, and the speedup is its ratio to the CUDA-Q rate. Supplementary Note~\ref{app:flat-allocation} discusses what repeated readout from shared trajectories means statistically.

The CUDA-Q reference produces 0.723 shots/s, from 1,333 to 1,344 shots per window. The rates of the three runs of each budget agree to within 2\%, so the error bars of Fig.~\ref{Fig:Fig4} are hidden by the markers. The largest ratio is 70,822.

\begin{figure*}[tbp]
 \centering
 \includegraphics[width=\textwidth]{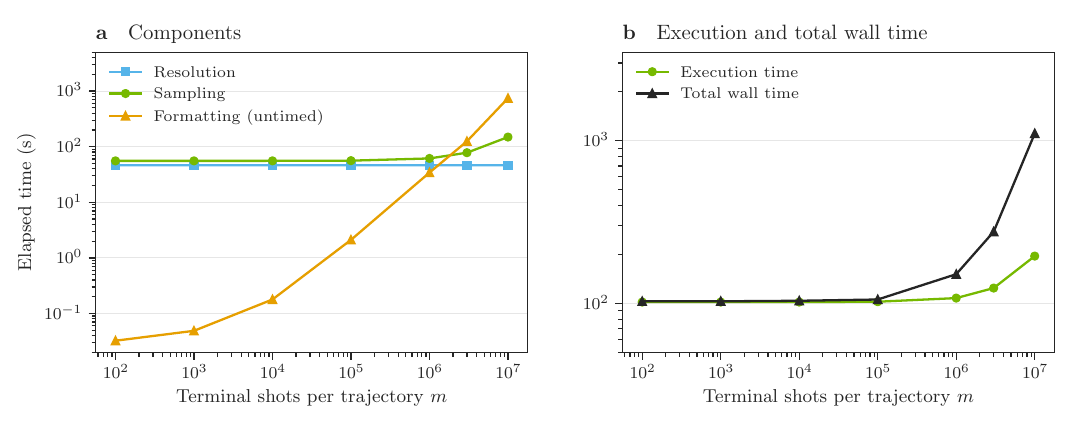}
 \caption{\textbf{Simulation and record-handling time for bulk generation.}
 Three runs per point on the $n=30$, $D=16$ amplitude-damping circuit, with 100 trajectories sampled $m$ times each on one H100 GPU. Points are means.
 \textbf{a,} Trajectory generation, sampling, and readout formatting. Only the first two count toward the execution time.
 \textbf{b,} Execution time and total wall time, which adds setup, formatting, and record handling.}
 \label{fig:supp-nonprop-timing}
\end{figure*}

Trajectory generation takes 46.3~s at every budget, and sampling grows from 55.5~s at $m=10^2$ to 149~s at $m=10^7$ (Fig.~\ref{fig:supp-nonprop-timing}). For $10^9$ shots the execution takes 195~s, or $5.1\times10^6$ shots/s. The whole run takes 1,106~s of wall time, most of it spent formatting the readout (744~s) and handling records outside the timer.

\FloatBarrier

\end{document}